\documentclass[12pt,dvipdfmx]{article}
\usepackage{amsmath,graphicx}

\font\mybb=msbm10 at 12pt
\def\bb#1{\hbox{\mybb#1}}
\def\Z {\bb{Z}}
\def\R {\bb{R}}
\font\mycc=msbm10 at 10pt
\def\cc#1{\hbox{\mycc#1}}

\def\mR {\cc{R}}
\font\mydd=msbm10 at 8pt
\def\dd#1{\hbox{\mydd#1}}

\def\sR {\dd{R}}

\def\cO {{\cal O}}

\def\tr{\mathop{\rm tr}\nolimits}

\def\diag{\mathop{\rm diag}\nolimits}
\def\mod{\mathop{\rm mod}\nolimits}

\makeatletter
\@addtoreset{equation}{section}
\@addtoreset{footnote}{section}
\makeatother

\newcommand{\be}{\begin{equation}}
\newcommand{\ee}{\end{equation}}
\newcommand{\wt}{\widetilde}
\newcommand{\wh}{\widehat}

\newcommand{\ra}{\rightarrow}

\newcommand{\nn}{\nonumber}
\newcommand{\half}{\frac{1}{2}}
\newcommand{\del}{\partial}

\begin{document}

\begin{flushright}
\hfill{YITP-16-139}
\end{flushright}
\begin{center}
\vspace{2ex}
{\Large {\bf 
Chern-Simons 5-form and Holographic Baryons
}}

\vspace*{5mm}
{\sc Pak Hang Chris Lau}$^{a,b}$\footnote{e-mail:
 {\tt pakhang.lau@yukawa.kyoto-u.ac.jp}}
~and~
{\sc Shigeki Sugimoto}$^{a,c}$\footnote{e-mail:
 {\tt sugimoto@yukawa.kyoto-u.ac.jp}}

\vspace*{4mm} 

\hspace{-0.5cm}
{\it {$^{a}$
Center for Gravitational Physics, Yukawa Institute for Theoretical
 Physics,\\ Kyoto University, Kyoto 606-8502, Japan
}}\\
{\it {$^{b}$
Center for Theoretical Physics, Massachusetts Institute of Technology,
\\ Cambridge, MA02139, USA
}}\\
{\it {$^{c}$
Kavli Institute for the Physics and Mathematics of the Universe (WPI),\\
 The University of Tokyo, Kashiwanoha, Kashiwa 277-8583, Japan}}\\ 

\end{center}

\vspace*{.3cm}
\begin{center}
{\bf Abstract}
\end{center}

In the top-down holographic model of QCD based on D4/D8-branes in type
IIA string theory and some of the bottom up models, the low energy
effective theory of mesons is described by a 5 dimensional
Yang-Mills-Chern-Simons theory in a certain curved background with two
boundaries. The 5 dimensional Chern-Simons term plays
a crucial role to reproduce the correct chiral anomaly in 4
dimensional massless QCD. However, there are some subtle ambiguities in
the definition of the Chern-Simons term for the cases with
topologically non-trivial gauge bundles, which include the configurations
with baryons.
In particular, for the cases with three flavors, it was
pointed out by Hata and Murata that the naive Chern-Simons term does not
lead to an important constraint on the baryon spectrum, which is needed
to pick out the correct baryon spectrum observed in nature.
In this paper, we propose a formulation of well-defined Chern-Simons
term which can be used for the cases with baryons,
and show that it recovers the correct baryon constraint as well as the
chiral anomaly in QCD.

\newpage

\tableofcontents

\section{Introduction}

The gauge/gravity duality provides a powerful method to study strongly
coupled gauge theories using theories with gravity \cite{Mald,GKP,Witt}.
One of its surprising features is that the space-time
dimensions of the gravity side is higher than that of the
corresponding gauge theory. For this reason this type of duality is
called holographic duality.
It has been applied to QCD and there have been a lot of successes in
revealing the properties of QCD and physics of hadrons.\footnote{
See \cite{Guijosa:2016upo} for a recent review.}
The holographic dual description of QCD (or QCD-like theory) is called
holographic QCD.
A common feature of the
holographic models is that the meson effective action is given as a 5
dimensional gauge theory embedded in a certain curved background.

In this paper, our main focus is on the 5 dimensional Chern-Simons
term\footnote{Here, the gauge field $A$ is a 1-form
and its field strength $F=dA+A\wedge A$ is a 2-form
that take values in the anti-Hermitian matrices.
We often omit the symbol ``$\wedge$'' for the wedge products of the
differential forms.
}
\begin{eqnarray}
 S_{\rm CS}=C\int_{M_5}\omega_5(A)\ ,
\label{CS}
\end{eqnarray}
where $C$ is a constant and $\omega_5(A)$ is the CS 5-form
that satisfies $d\omega_5(A)=\tr(F^3)$. The explicit form of the CS
5-form is
\begin{eqnarray}
\omega_5(A)\equiv\tr\left( AF^2-\half A^3F+\frac{1}{10}A^5 \right)
=\tr\left( AdAdA +\frac{3}{2}A^3dA +\frac{3}{5}A^5 \right) \ .
\label{CS5}
\end{eqnarray}
It appears in the meson effective action in the top-down holographic
model of QCD proposed in \cite{SS}\footnote{
See \cite{Rebhan:2014rxa} for a review.} and some of the bottom-up
models (See, e.g., \cite{Son:2003et,DH,Pomarol:2008aa,Domokos:2009cq}).
In these models, the effective theory of mesons is described by a 5
dimensional $U(N_f)$ Yang-Mills-Chern-Simons (YM-CS) action
on a curved space-time $M_5$, where $N_f$ is the number of massless
quarks, and the coefficient of the CS-term is related to the number of
color $N_c$ by
\begin{eqnarray}
C=\frac{i N_c}{24\pi^2}\ .
\end{eqnarray}
The normalizable modes of the 5 dimensional $U(N_f)$ gauge field $A$
correspond to the degrees of freedom of a tower of vector and axial
vector mesons (such as rho meson, omega meson, $a_1$ meson, etc.) as
well as the massless pions.\footnote{
In this paper, we consider the cases with massless quarks.
See \cite{quarkmass} for the proposals to include quark masses.
}
It has been shown that the masses as well as coupling constants for
low-lying mesons read off from the 5 dimensional YM-CS theory
turn out to be in reasonably good agreement with the experimental data
and provides some predictions for the unknown
parameters.

The CS term plays crucial roles in many aspects in holographic
QCD. First of all, the chiral anomaly in QCD is correctly reproduced due
to the CS term. In fact, the 5 dimensional expression of the WZW term
in QCD \cite{Wess:1971yu,Witt2,Kaymakcalan:1983qq} has a direct physical
interpretation in terms of the 5 dimensional CS term in
holographic QCD \cite{SS}.
Furthermore,
some of the decay modes of the omega meson ($\omega\ra\pi^0\gamma$ and
$\omega\ra\pi^0\pi^+\pi^-$) are induced by terms generated from the CS
term. Surprisingly, the structure of the interaction terms for these
decay modes predicted by holographic QCD agrees with that of the
Gell-Mann--Sharp--Wagner model \cite{GSW}, which is a phenomenological
model proposed to reproduce the experimental data of the omega meson
decay \cite{SS2} (See also \cite{Pomarol:2008aa}.).
The CS term is also important in the analysis of baryons.
Due to the CS term, it can be shown that the baryon number
is equal to the instanton number defined on a time slice \cite{SS}.
When the vector (and axial-vector) mesons are integrated out,
the 5 dimensional YM-CS action reduces to the action
of the Skyrme model \cite{SS,SS2}.
The Skyrme model was proposed by Skyrme to describe
baryons as topological solitons called Skyrmion \cite{Skyrme}.
The pion field in the soliton has a non-trivial winding number
representing an element of the homotopy group
$\pi_3(U(N_f))\simeq \Z$. The relation between the instanton number for
the 5 dimensional gauge field and the winding number carried by the pion
field is precisely that proposed by Atiyah and Manton \cite{AM} in an
attempt to obtain approximate Skyrmion solutions
by using instanton solutions.

However, there are some subtle ambiguities in the definition of the CS
term. In the explicit expression of the CS term in (\ref{CS}) with
(\ref{CS5}), we have implicitly assumed that the gauge field $A$ is a
globally well-defined 1-form on the 5 dimensional space-time $M_5$.
This is, however, not always possible when the gauge configuration with
a given boundary condition is topologically non-trivial,
including the cases with baryons. In such
cases, it is necessary to cover the 5 dimensional space-time $M_5$ by
multiple patches on which the gauge field is well-defined. One might
naively think that the CS term can be defined as just a sum of the CS
term defined on each patch. However, this approach doesn't work, because
it depends on the choice of the gauge, and some additional terms are
needed to make it well-defined.
Related to this issue, a problem was pointed out by Hata and Murata in
\cite{HM}. They tried to analyze the spectrum of baryons in the case
with $N_f=3$, generalizing the analysis for $N_f=2$ in \cite{HSSY}, and
claimed that a constraint needed to get the correct baryon spectrum
(see (\ref{constraint})) cannot be obtained by using the naive CS
term. They proposed a new CS term that gives the correct
constraint, but it does not reproduce the chiral anomaly of QCD.
Our main goal is to propose a well-defined CS term 
that solves all these problems.

The paper is organized as follows. We start with reviewing the
problems in more detail while fixing our notation in section
\ref{Puzzle}.
Our proposal for the well-defined CS term is given in section
\ref{Proposal}. In section \ref{Baryon_spec}, we revisit the analysis of
the effective action for the collective coordinates
of the soliton solution representing baryons
and show that the correct
constraint is obtained from the new CS term. Section \ref{Conclusion}
gives a summary and outlook.

\section{Puzzle}
\label{Puzzle}

\subsection{The model}
\label{model}

Our starting point is the 5 dimensional $U(N_f)$ YM-CS action given by
\be
S_{\rm 5dim}=S_{\rm YM}+S_{\rm CS} \,,
\label{S5dim}
\ee
with $S_{\rm CS}$ as defined in (\ref{CS}) and the kinetic term
for the gauge field
\begin{align}
S_{\rm YM}&=-\frac{\kappa}{2}\int_{M_5}\tr (F\wedge *F)\, ,
\label{YM}
\end{align}
where $\kappa$ is a constant and
$*$ is the Hodge star in 5 dimensional space-time $M_5$.
Although the details of the metric on $M_5$ is not important in our main
purpose, we use the following form of the metric for explicit
calculations:
\be
ds^2=4(k(z)\wt k(z) \eta_{\mu\nu}dx^\mu dx^\nu+ \wt k(z)^2 dz^2) \, ,
\label{eqn:metric}
\ee
where $x^{\mu}$ ($\mu=0,1,2,3$) are the coordinates for the 4
dimensional Minkowski space-time and $z$ is the coordinate for the fifth
direction. Then, the Hodge dual of the field strength 2-form $F$ is
\be
*F=- \frac{k(z)}{3}F^\mu_{~z} \epsilon_{\mu\nu\rho\sigma}
 dx^{\nu} dx^{\rho} dx^{\sigma}
+\frac{\wt k(z)}{2} F^{\mu\nu} \epsilon_{\mu\nu\rho\sigma}
dx^{\rho}dx^{\sigma} dz\ ,
\label{eqn:starF}
\ee
where $\epsilon_{\mu\nu\rho\sigma}$ is the totally antisymmetric tensor
in 4 dimensional Minkowski space with $\epsilon_{0123}=+1$, and the
Lorentz indices are raised and lowered by the Minkowski metric
 $(\eta_{\mu\nu})=(\eta^{\mu\nu}) = \diag(-1,1,1,1)$.
Then, the YM action (\ref{YM}) is written as
\be
S_{\rm YM}=\kappa \int d^4xdz \tr \left(\frac{1}{2}\, \wt k(z)
 F_{\mu\nu} F^{\mu\nu} + k(z)
 F_{\mu z}F^{\mu}_{~ z}\right)\, .
\label{eqn:act_met}
\ee
The meson effective action in \cite{SS}
is given by (\ref{S5dim}) with $\wt k(z)=(1+z^2)^{-1/3}$ and $k(z)=1+z^2$.

The boundary of $M_5$ is a disjoint union of the 4 dimensional
edges at $z\ra +\infty$ and $z\ra -\infty$:\footnote{
Note that the asymptotic region at $|x^\mu|\ra\infty$ is not regarded as
the boundary. In order to avoid confusion, we compactify the $x^\mu$
directions in the following discussion.
}
\begin{eqnarray}
 \del M_5=M_4^{(+\infty)}\cup(-M_4^{(-\infty)})\ ,
\end{eqnarray}
where $M_4^{(\pm\infty)}\equiv M_5|_{z\ra\pm\infty}$ and
the minus sign in front of $M_4^{(-\infty)}$ means the orientation
is reversed.
The boundary values of the gauge field pulled back on
$M_4^{(\pm\infty)}$, denoted as $A|_{z\ra\pm\infty}(=
\lim_{z\ra\pm\infty} A_\mu dx^\mu)$,
are interpreted as the external gauge fields associated with the chiral
symmetry $U(N_f)_L\times U(N_f)_R$ in QCD.\footnote{The axial $U(1)$
subgroup of $U(N_f)_L\times U(N_f)_R$ is anomalous. This anomaly can
also be seen in string theory as discussed in \cite{SS}, but we won't
discuss it here.}
More precisely, we set $\wh A_\pm = A|_{z\ra\pm\infty}$, where
 $\wh A_+$ and $\wh A_-$ are the external gauge fields
associated with $U(N_f)_R$ and $U(N_f)_L$, respectively.
Because the gauge field at the boundary is fixed, the gauge symmetry
of the system consists of the gauge transformation that acts trivially
at the boundaries.
The gauge transformation at $z\ra\pm\infty$ corresponds to that of the
chiral symmetry. Note that the CS term (\ref{CS}) is not invariant under
the gauge transformation that acts non-trivially at the boundary.
In fact, the infinitesimal gauge transformation of the CS term
with $\delta_\Lambda A=d\Lambda +[A,\Lambda]\equiv D_A\Lambda$ is
\begin{eqnarray}
 \delta_\Lambda S_{\rm CS}=C
\left(\int_{M_4^{(+\infty)}}\omega_4^1(\wh\Lambda_+,\wh A_+)
-\int_{M_4^{(-\infty)}}\omega_4^1(\wh\Lambda_-,\wh A_-)
\right)\ ,
\label{deltaSCS}
\end{eqnarray}
where 
$\wh\Lambda_\pm\equiv \Lambda|_{z\ra\pm\infty}$
and
\begin{eqnarray}
 \omega_4^1(\Lambda,A)\equiv\tr\left(\Lambda d\left(
AdA+\half A^3
\right)\right)\ .
\end{eqnarray}
Here, we have used the formula
\begin{eqnarray}
\delta_\Lambda\omega_5(A)=d\omega_4^1(\Lambda,A)+\cO(\Lambda^2)\ ,
\end{eqnarray}
and the Stokes' theorem.\footnote{See Appendix \ref{notations}
for our notations and useful formulae.}
(\ref{deltaSCS}) precisely agrees with the chiral anomaly in
QCD.\footnote{See, e.g., a textbook \cite{Nair:2005iw} for a review
of anomaly.}

\subsection{Problems of the CS term}
\label{problemCS}

In order to illustrate the problem clearly, let us compactify the time
and $x^{1\sim 3}$ directions, and consider the case that the topology of
the space-time is equivalent to
\begin{eqnarray}
 M_5\simeq S^1\times S^3\times\R  \ ,
\end{eqnarray}
where $S^1$ is the compactified time direction,
$S^3$ is the compactified $x^{1\sim 3}$ directions
and $\R$ is the $z$ direction.\footnote{
To be more precise, we add the boundary points $\{z\ra\pm\infty\}$
to $\mR$ and treat the $z$ direction as a closed interval
$I=[-\infty,+\infty]$.
}
As shown in \cite{SS}, the baryon number $n_B$ is given by the instanton
number on a time slice (see also section \ref{eom-current} for a
derivation):
\begin{eqnarray}
n_B=\frac{1}{8\pi^2}\int_{S^3\times\sR}\tr(F^2)\ .
\label{nB}
\end{eqnarray}
When the gauge field $A$ is a globally well-defined 1-form on $M_5$,
using the formula
\begin{eqnarray}
 \tr(F^2)=d\omega_3(A) \,,
\end{eqnarray}
with the CS 3-from
\begin{eqnarray}
 \omega_3(A)\equiv\tr\left(AF-\frac{1}{3} A^3
\right)
=\tr\left(AdA+\frac{2}{3} A^3 \,,
\right)
\end{eqnarray}
and the Stokes' theorem, (\ref{nB}) can be rewritten as
\begin{eqnarray}
n_B=\frac{1}{8\pi^2}\left(\int_{S^3}\omega_3(A)|_{z\ra+\infty}
-\int_{S^3}\omega_3(A)|_{z\ra -\infty}
\right)
\ .
\label{nB2}
\end{eqnarray}
This expression inevitably vanishes if we impose the boundary condition
$A|_{z\ra\pm\infty}=0$. Therefore, if we adopt
the identification $\wh A_\pm=A|_{z\ra\pm\infty}$ in the previous subsection,
the globally well-defined gauge field $A$ can describe only the $n_B=0$
sector of the gauge configuration,
when the external gauge fields $\wh A_\pm$ are turned off.
This is clearly restricting the gauge configurations too much.
As usual in gauge theory, we should include the gauge configurations
defined on topologically non-trivial gauge bundles.

\begin{figure}[ht]
\centering
\includegraphics[width=6cm]{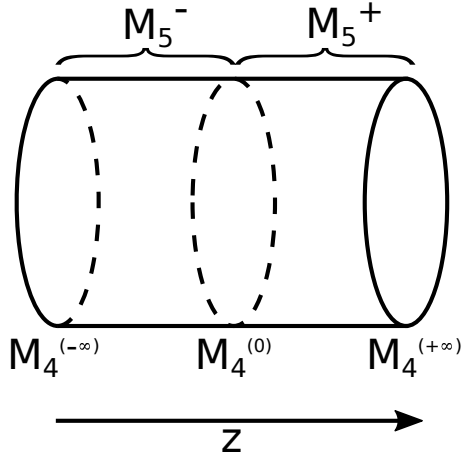}
\caption{The 5 dimensional space-time $M_5$}
\label{fig:M5_mani}
\end{figure}

In order to describe gauge configurations with non-zero baryon
number, we cover the space-time manifold $M_5$ with two patches as
\begin{eqnarray}
 M_5=M_5^-\cup M_5^+\ ,
\end{eqnarray}
where $M_5^\pm$ are chosen to be
$M_5^\pm\equiv \{(x^\mu,z)\in M_5\,|\, \pm z> -\epsilon\}$
with a small positive parameter $\epsilon$. The intersection of the two
patches is
\begin{eqnarray}
M_5^-\cap M_5^+\simeq M_4^{(0)}\times(-\epsilon,+\epsilon)\ ,
\end{eqnarray}
where $M_4^{(0)}\equiv\{(x^\mu,z)\in M_5\,|\,z=0\}\simeq S^1\times S^3$.
In the following, we understand $\epsilon$ as an infinitesimal parameter
and take the limit $\epsilon\ra 0$ at the end of the calculations.
The picture in the
 $\epsilon\ra 0$ limit is depicted in Figure \ref{fig:M5_mani}.
The gauge configuration is defined by the gauge field $A_\pm$ defined
on each patch $M_5^\pm$
\footnote{For simplicity, we have assumed here that
$A_\pm$ are well-defined $U(N_f)$ valued 1-forms on $M_5^\pm$.
This is always the case for a static gauge configuration and a
small perturbation around it,
because the gauge bundle over $S^3$ is trivial due to
$\pi_2(U(N_f))\simeq 0$.
A counterexample is a gauge configuration
with non-zero instanton number defined on $S^1\times S^3$, 
which looks like a baryon configuration with the time and $z$ directions
interchanged. General gauge configurations may be described by
introducing more patches to have good covering of $M_5$,
though we won't discuss the details here.
}
and connected by the gluing condition on the intersection as
\begin{eqnarray}
 A_+=A_-^h\equiv hA_-h^{-1}+hdh^{-1}\ ,~~~({\rm on}~M_5^-\cap M_5^+)
\label{glue}
\end{eqnarray}
where $h$ is a $U(N_f)$ valued function defined on the intersection
$M_5^-\cap M_5^+$.
The external gauge fields $\wh A_\pm$ are now related to the boundary
values of the gauge fields $A_\pm$ as
\begin{eqnarray}
 \wh A_\pm\equiv A_\pm|_{z\ra\pm\infty} \ .
\label{Abdry0}
\end{eqnarray}
The gauge transformation is given by
\begin{eqnarray}
A_\pm\ra A_\pm^{g_\pm}\equiv g_{\pm}A_\pm g_{\pm}^{-1}
+g_{\pm}dg_{\pm}^{-1}\ ,~~~ 
h\ra g_+ h g_-^{-1}\ ,
\label{gauge}
\end{eqnarray}
where $g_\pm$ are $U(N_f)$ valued functions on $M_5^\pm$.
The boundary values of the gauge functions
$\wh g_\pm\equiv g_\pm|_{z\ra\pm\infty}$ correspond to those of
the (gauged) chiral symmetry as
$(\wh g_-,\wh g_+)\in U(N_f)_L\times U(N_f)_R$.

In this setup, it is possible to have gauge configurations with
non-zero baryon number.
In fact,  (\ref{nB}) gives
\begin{eqnarray}
n_B=
\frac{1}{24\pi^2}\int_{S^3}\tr((hdh^{-1})^3)|_{z=0}
\label{nB3}
\end{eqnarray}
for the case with $\wh A_\pm=0$. The baryon number (\ref{nB3})
is equivalent to the winding number given as an element of
$\pi_3(U(N_f))\simeq\Z$ represented by the $U(N_f)$ valued function
$h|_{z=0}$ restricted at a time slice.

The question now is how to define the CS term in this setup.
While the CS term is supposed to give the correct chiral anomaly,
we should make sure that it is invariant
(up to a $2\pi$ shift) under the gauge transformations
with $\wh g_\pm=1$ that act trivially at the boundary.
One can immediately see that a naive expression like
\begin{eqnarray}
C\left(\int_{M_5^-}\omega_5(A_-)+
\int_{M_5^+}\omega_5(A_+)
\right) \ ,
\label{naiveCS}
\end{eqnarray}
does not work. This is one of the reasons that the naive CS term
has to be modified.

Another approach is to insist on a globally well-defined gauge field
$A$, and modify the relation between the boundary values of the gauge
field and the external gauge field associated with the chiral symmetry.
This can be achieved from the above description by the gauge
transformation (\ref{gauge}) with $g_\pm=h_\pm$ satisfying
$h_+hh_-^{-1}=1$ on $M_5^-\cap M_5^+$. Then, the gauge field $A$
defined as
\begin{eqnarray}
A\equiv A_\pm^{h_\pm}~~{\rm on}~~M_5^\pm
\label{globalA}
\end{eqnarray}
is a globally well-defined 1-form on $M_5$, because the gluing condition
(\ref{glue}) implies $A_+^{h_+}=A_-^{h_-}$ on the intersection
$M_5^-\cap M_5^+$.
In this case, because of the relation (\ref{Abdry0}),
the boundary values of the gauge field $A$ are not equal
to the external gauge fields $\wh A_\pm$, but related by the gauge
transformation as
\begin{eqnarray}
 A|_{z\ra\pm\infty} =\wh A_\pm^{\,\wh h_\pm}\ ,
\label{Abdry}
\end{eqnarray}
where $\wh h_\pm\equiv h_\pm|_{z\ra\pm\infty}$.
It is important to note that a gauge configuration is specified
by the pair $(A,\wh h_\pm)$. Two gauge configurations
with the same gauge field $(A,\wh h_\pm)$ and $(A,\wh h_\pm')$ can be
physically inequivalent when $\wh h_\pm$ and $\wh h_\pm'$ are different.

It is easy to see that, with the identification (\ref{Abdry}), the
expressions for the baryon number (\ref{nB2}) and (\ref{nB3}) are
identical.
When the external gauge fields are turned off, the boundary values
of the gauge field are given by
 $A|_{z\ra\pm\infty}=\wh h_\pm d\wh h_\pm^{-1}$
and the baryon number (\ref{nB2}) is given by the difference of the
winding number carried by $\wh h_+$ and $\wh h_-$ as
\begin{eqnarray}
n_B=-\frac{1}{24\pi^2}\int_{S^3}
\left(
\tr((\wh h_+d\wh h_+^{-1})^3)-\tr((\wh h_-d\wh h_-^{-1})^3)
\right)\ .
\label{nB4}
\end{eqnarray}
Therefore, for the gauge configurations with non-zero baryon number,
$\wh h_\pm$ cannot be trivial and the gauge field $A$ does not vanish at
the boundaries.

One might think that the naive CS term (\ref{CS}) can be used for this
globally well-defined gauge field $A$. 
However, this CS term depends on the choice of the gauge,
since (\ref{CS}) is not invariant under the
gauge transformation that changes the boundary values.
To see this, consider a gauge transformation
\begin{eqnarray}
 A\ra A^g\ ,~~~\wh h_\pm\ra (g \wh h_\pm)|_{z\ra\pm\infty}\ ,
\label{Ahtr}
\end{eqnarray}
with a $U(N_f)$ valued function $g$ on $M_5$. This gauge transformation
does not act on the external gauge fields $\wh A_\pm$ and hence the
gauge configurations $(A,\wh h_\pm)$ and $(A^g, g\wh h_\pm)$ are
physically equivalent.
The problem is that $\omega_5(A)$ and $\omega_5(A^g)$
are not equal (see (\ref{gauge5})) and it is not clear 
which one we should use.
Moreover, the naive CS term (\ref{CS}) does not reproduce
the expression (\ref{deltaSCS}) for the chiral anomaly.
Because of the boundary condition (\ref{Abdry}), 
the relation between the boundary values of the gauge
function $g$ in the gauge transformation $A\ra A^g$ and the gauge
function
for the gauged chiral symmetry $\wh g_\pm$ is modified as
\begin{eqnarray}
 \wh g_\pm=(\wh h_\pm^{-1}g \wh h_\pm)|_{z\ra\pm\infty}\ .
\label{chtr}
\end{eqnarray}
Then, the transformation $(A,\wh h)\ra (A^g,\wh h)$ induces
$\wh A_\pm\ra\wh A_\pm^{\,\wh g_\pm}$ as desired.
For the infinitesimal gauge transformation with $g\simeq 1-\Lambda$ and
$\wh g_\pm\simeq 1-\wh\Lambda_\pm$, (\ref{chtr}) gives
$\wh\Lambda_\pm=(\wh h_\pm^{-1}\Lambda \wh h_\pm)|_{z\ra\pm\infty}$
and hence the infinitesimal gauge
transformation of the naive CS term (\ref{CS}) is
\begin{eqnarray}
 \delta_\Lambda S_{\rm CS}=C
\left(\int_{M_4^{(+\infty)}}
\omega_4^1(\wh h_+\wh\Lambda_+\wh h_+^{-1},\wh A_+^{\,\wh h_+})
-\int_{M_4^{(-\infty)}}
\omega_4^1(\wh h_-\wh\Lambda_-\wh h_-^{-1},\wh A_-^{\,\wh h_-})
\right)\ ,
\end{eqnarray}
which does not agree with (\ref{deltaSCS}) in general.

In addition to these issues, there is a more practical problem of the CS
term pointed out by Hata and Murata in \cite{HM}. 
They studied the spectrum of baryons in holographic QCD with $N_f=3$.
The analysis is similar to that for the 3-flavour Skyrme model.
In Skyrme model, baryons are represented as topological solitons
called Skyrmions in a theory of pion. 
There are collective coordinates corresponding to the
$SU(3)$ rotation (for $N_f=3$) of the Skyrmion solution, which are
denoted by $a\in SU(3)$. (See section \ref{skyrme}.)
It has been shown that the WZW term gives
\begin{eqnarray}
 S_{\rm WZW}=-i\frac{N_c n_B}{\sqrt{3}}\int dt
\tr(t_8 a^{-1} \del_t a)\ ,
\label{WZW}
\end{eqnarray}
which leads to a constraint
\begin{eqnarray}
 \psi(a\, e^{i t_8\theta})
=\psi(a)\exp\left(i\frac{N_c n_B}{2\sqrt{3}}\theta\right)\ ,
\label{constraint}
\end{eqnarray}
on the wave function $\psi(a)$ for the quantum mechanics of the
collective
coordinates \cite{Witten:1983tx,Guadagnini:1983uv,
Chemtob:1985ar,Jain:1984gp,Manohar:1984ys,BLR}\footnote{
See also a textbook \cite{Nair:2005iw} for a review.}.
Here,
\begin{eqnarray}
t_8\equiv\frac{1}{2\sqrt{3}}
\left(
\begin{array}{ccc}
 1& & \\
 &1 & \\
 & &-2 \\
\end{array}
\right)
\end{eqnarray}
is the 8th generator of the $SU(3)$ algebra.
This constraint is crucial to obtain the baryon spectrum
consistent with the experiments.
Since the WZW term can be derived from the CS term in holographic QCD
\cite{SS}, it is natural to expect that the CS term plays a similar
role. However, it was claimed that the contribution from the CS term
vanishes and the constraint (\ref{constraint}) cannot be
reproduced, by using the naive CS term (\ref{CS}) in a certain
gauge.
In order to get the correct constraint (\ref{constraint}),
they proposed to use the CS term of the form
\begin{eqnarray}
 S_{\rm CS}^{\rm HM}=C\int_{M_6}\tr(F^3)\ ,
\label{HMCS}
\end{eqnarray}
where $M_6$ is a 6 dimensional manifold with $\del M_6=M_5$.
Although they succeeded in recovering the correct constraint by
using this new CS term, it is also problematic.
First, as emphasised above, $M_5$ has boundaries and
the meaning of ``$\del M_6=M_5$'' is not clear, because
$\del M_5=\emptyset$ is
a necessary condition to have such $M_6$.
Furthermore,
this term is manifestly gauge invariant and it does not recover
the chiral anomaly (\ref{deltaSCS}).

\section{Proposal}
\label{Proposal}

In this section, we propose a new CS term that solves all the problems
discussed in the previous section.

\subsection{Proposal for the CS-term}
\label{proposal1}

Using the notation introduced in section \ref{problemCS},
our proposal for the CS term is given by
\begin{eqnarray}
S_{\rm CS}^{\rm new}\equiv
C\left(
\int_{M_5^-}\omega_5(A_-)
+\int_{M_5^+}\omega_5(A_+)
+\frac{1}{10}\int_{N_5^{(0)}}\tr\left((\wt hd\wt h^{-1})^5\right)
+\int_{M_4^{(0)}}\alpha_4(dh^{-1}h,A_-)
\right)\ ,
\label{eqn:newCS}
\end{eqnarray}
where $N_5^{(0)}$ is a 5 dimensional manifold satisfying
$\del N_5^{(0)}=M_4^{(0)}$, $\wt h$ is a $U(N_f)$ valued function on
 $N_5^{(0)}$ satisfying $\wt h|_{\del N_5^{(0)}}=h$, and
\begin{eqnarray}
 \alpha_4(V,A)&\equiv&
\half\tr\left(V(A^3-AF-FA)+\half VAVA+V^3A\right)
\nn\\
&=&
-\half\tr\left(V(AdA+dAA+A^3)-\half VAVA-V^3A\right)
\ .
\end{eqnarray}
Useful formulae for the CS 5-form $\omega_5(A)$ and the 4-form
$\alpha_4(V,A)$ can be found in Appendix \ref{CS5form}.
Note that the last term in (\ref{eqn:newCS}) can be replaced with
\begin{eqnarray}
-C\int_{M_4^{(0)}}\alpha_4(dh h^{-1},A_+)\ ,
\end{eqnarray}
using (\ref{alpha4gauge}). The third and fourth terms in
(\ref{eqn:newCS}) are added to the naive expression (\ref{naiveCS}).
The motivation for adding these terms will soon become clear.

A few comments are in order.
In (\ref{eqn:newCS}), we have assumed the existence of
$N_5^{(0)}$ and $\wt h$.\footnote{
For a generic choice of $M_4^{(0)}$ and $h$,
the existence of $N_5^{(0)}$ and
$\wt h$ is not guaranteed. For example, for
$M_4^{(0)}={\bf CP}^2$, which is known to be a non-trivial element
of the cobordism group for oriented closed 4-manifolds, $N_5^{(0)}$
does not exist.
On the other hand, when $M_4^{(0)}= S^1\times M_3$ with $M_3$ being a closed
oriented 3-manifold $M_3$, there always exists a 4-manifold $N_4$
satisfying $\del N_4=M_3$ and $N_5^{(0)}$ can be either $D\times M_3$ or
$S^1\times N_4$.
If $h$ is topologically non-trivial on $M_3$, like the examples
with $n_B\ne 0$ considered in section \ref{problemCS},
we should choose $N_5^{(0)}=D\times M_3$ so that $\wt h$ defined on
$N_5^{(0)}$ can be found. However, if $h$ has a non-trivial winding
number as a map from $S^1$ to $U(N_f)$ at each point in $M_3$,
this is not possible. For this reason, we consider the cases that $h$
does not wind around a non-trivial 1-cycle in $U(N_f)$ along the $S^1$
direction.
For the case of $M_4^{(0)}\simeq S^4$, we can choose
$N_5^{(0)}$ to be a 5 dimensional ball and then $\wt h$ always
exists for $N_f\ge 3$, because $\pi_4(U(N_f))$ is trivial.
}
For the case with $M_4^{(0)}\simeq S^1\times S^3$ and $h\in SU(N_f)$,
which is the case of our main interest,
one can choose $N_5^{(0)}$ to be $N_5^{(0)}\simeq D\times S^3$, where
$D$ is a disk satisfying $\del D=S^1$, and then $\wt h$ exists
because the image of $h$, as a map from $S^1$ to $SU(N_f)$ at
each point in $S^3$, is contractible in $SU(N_f)$.
The choice of $N_5^{(0)}$ and $\wt h$ does not matter, due to the standard
argument for the WZW term \cite{Witt2}. 

This new CS term has the following desired properties:
\begin{enumerate}
\item It reduces to (\ref{CS}) when $h$ is topologically trivial.
\item It is invariant (up to a $2\pi\Z$ shift) under the gauge
      transformation (\ref{gauge}) with $g_\pm|_{z\ra\pm\infty}\ra 1$.
\item It reproduces the correct chiral anomaly in QCD (\ref{deltaSCS})
with the identification $\wh A_\pm=A_{\pm}|_{z\ra\pm\infty}$
and $\wh g_\pm =e^{-\wh\Lambda_\pm}=g_\pm|_{z\ra\pm\infty}$.
\item It reduces to the Hata-Murata's proposal (\ref{HMCS})
when $M_5$ does not have boundaries, ({\it i.e.}
$M_4^{(\pm\infty)}=\emptyset$), and there exists a 6 dimensional
manifold $M_6$ such that $\del M_6=M_5$ and
$M_6=M_6^+\cup M_6^-$ with
$M_6^+\cap M_6^-\simeq N_5^{(0)}\times (-\epsilon,\epsilon)$
and $\del M_6^\pm\simeq M_5^\pm\cup(\pm N_5^{(0)})$.
(see Figure \ref{fig:M6_mani} for the picture in the limit
 $\epsilon\ra 0$.)
\begin{figure}[ht]
\centering
\includegraphics[width=6cm]{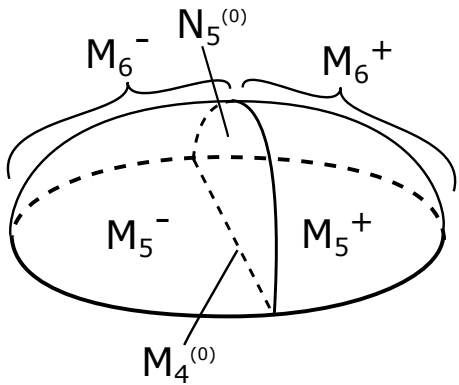}
\caption{The 6 dimensional space-time $M_6$.}
\label{fig:M6_mani}
\end{figure}
\end{enumerate}

Let us show these properties one by one.
\begin{enumerate}
\item 
When $h$ is topologically trivial, {\it i.e.} $h$ can be continuously
deformed to $h=1$, there exists a $U(N_f)$ valued
function $\wt h$ on $M_5^-$ such that $\wt h=h$ on the intersection
$M_5^-\cap M_5^+$ and satisfy the boundary condition
$\wt h|_{z\ra -\infty}\ra 1$. Then, we can obtain a globally
well-defined 1-form $A$ on $M_5$ by defining
\begin{eqnarray}
A\equiv\left\{
\begin{array}{lc}
A_-^{\wt h}&\mbox{(on~$M_5^-$)}\\
A_+&\mbox{(on~$M_5^+$)}
\end{array} \,.
\right.
\label{Agl}
\end{eqnarray}
We choose $N_5^{(0)}=M_5^-\cup N_5^{(-\infty)}$, where
$N_5^{(-\infty)}$ is a 5 dimensional manifold with
$\del N_5^{(-\infty)}=M_4^{(-\infty)}$, and
extend $\wt h$ to $N_5^{(0)}$ by setting $\wt h|_{N_5^{(-\infty)}}=1$.
Then, we obtain
\begin{align}
S_{\rm CS}^{\rm new}
&= C\left(\int_{M_5^-}\omega_5(A_-)
+\int_{M_5^+}\omega_5(A_+) + \int_{M_5^-}
 \left[\frac{1}{10}\tr\left((\wt hd\wt h^{-1})^5\right)
+d\alpha_4(d\wt{h}^{-1}\wt{h},A_-) \right] \right) \nn\\
&=C\left(\int_{M_5^-}\omega_5(A_-^{\wt h}) +
 \int_{M_5^+}\omega_5(A_+)\right)\nn\\
&=C\int_{M_5}\omega_5(A)\ ,
\label{new2old}
\end{align}
where (\ref{gauge5}) is used.
\item
Under the gauge transformation (\ref{gauge}), the CS
term (\ref{eqn:newCS}) is transformed as
\begin{eqnarray}
&& S_{\rm CS}^{\rm new}
\nn\\
&\ra&
C\left(
\int_{M_5^-}\omega_5(A_-^{g_-})+
\int_{M_5^+}\omega_5(A_+^{g_+})
+\frac{1}{10}\int_{N_5^{(0)}}\tr\left((\wt h'd\wt h'^{-1})^5\right)
+\int_{M_4^{(0)}}\alpha_4(dh'^{-1}h',A_-^{g_-})
\right)\,, \nn \\
\label{newCStr}
\end{eqnarray}
where $h'\equiv g_+ h g_-^{-1}$ and $\wt h'$ are $U(N_f)$ valued
functions on $M_5^-\cap M_5^+$ and
$N_5^{(0)}$, respectively, satisfying $\wt h'|_{\del N_5^{(0)}}=h'|_{z=0}$.
Note that since $g_\pm|_{z=0}$ are topologically trivial
due to the boundary conditions $g_\pm|_{z\ra\pm\infty}\ra 1$, 
there exist $U(N_f)$ valued functions $\wt g_\pm$ on $N_5^{(0)}$
satisfying $\wt g_\pm|_{\del N_5^{(0)}}=g_\pm|_{z=0}$ and
$\wt h'$ can be constructed by $\wt h'=\wt g_+\wt h\wt g_-^{-1}$.
Then, using (\ref{gauge5}),
(\ref{alpha4gauge2}), (\ref{eqn:v5-2}) and (\ref{v5}),
one can show that (\ref{newCStr}) is equal to 
\begin{eqnarray}
&&C\left(
\int_{M_5^-}\omega_5(A_-)
+\int_{M_5^+}\omega_5(A_+)
+\frac{1}{10}\int_{N_5^{(0)}}\tr\left((\wt hd\wt h^{-1})^5\right)
+\int_{M_4^{(0)}}\alpha_4(dh^{-1}h,A_-)
\right)\ ,\nn\\
&&+\frac{C}{10}\left(\int_{M_5^+}\tr\left(G_+^5\right)
+\int_{N_5^{(0)}}\tr\left(\wt G_+^5\right)\right)
+\frac{C}{10}\left(\int_{M_5^-}\tr\left(G_-^5\right)
-\int_{N_5^{(0)}}\tr\left(\wt G_-^5\right)\right)\ ,
\label{newCStr2}
\end{eqnarray}
where $G_\pm\equiv dg_\pm^{-1}g_\pm$ and
 $\wt G_\pm\equiv d\wt g_\pm^{-1}\wt g_\pm$.
The first line is $S^{\rm new}_{\rm CS}$ defined in (\ref{eqn:newCS}).
The second line can be omitted because it takes value in
$2\pi \Z$.
\item 
Here, we consider the infinitesimal gauge transformation with
$\wh g_\pm \simeq 1-\Lambda_\pm$.\footnote{
See section \ref{other} for the finite transformation.
}
In this case, $g_\pm|_{z=0}$ is again topologically trivial
and it suffices to show property 3 for the cases
with $g_\pm=1$ on $M_5^-\cap M_5^+$, because of the
property 2 shown above.
Then, since the third and forth terms in (\ref{eqn:newCS})
do not change under the gauge transformation,
the proof of (\ref{deltaSCS}) is the same as that reviewed
in section \ref{model}.
\item
Using the relations $\del M_6^\pm=M_5^\pm\cup(\pm N_5^{(0)})$
and the Stokes' theorem, we obtain
\begin{eqnarray}
 S_{\rm CS}^{\rm HM}
&=&
C\left(
\int_{M_6^-}d\omega_5(A_-)+\int_{M_6^+}d\omega_5(A_+)
\right)
\nn\\
&=&
C\left(
\int_{M_5^-}\omega_5(A_-)+\int_{M_5^+}\omega_5(A_+)
+\int_{N_5^{(0)}}\left(\omega_5(A_+)-\omega_5(A_-)\right)
\right)\ .
\end{eqnarray}
Now, $A_+$ and $A_-$ are related by $A_+=A_-^{\wt h}$
on $M_6^-\cap M_6^+\simeq N_5^{(0)}\times (-\epsilon,+\epsilon)$.
Then, it is easy to check, using (\ref{gauge5}),
\begin{eqnarray}
\int_{N_5^{(0)}}\left(\omega_5(A_+)-\omega_5(A_-)\right)
=\int_{N_5^{(0)}}\frac{1}{10}\tr\left((\wt h d\wt h^{-1})^5\right) 
+\int_{\del N_5^{(0)}}\alpha_4(d\wt h^{-1}\wt h,A_-)\ ,
\end{eqnarray}
which shows that $S_{\rm CS}^{\rm HM}$ (\ref{HMCS}) agrees with
$S_{\rm CS}^{\rm new}$ (\ref{eqn:newCS}).
\end{enumerate}

\subsection{Other useful expressions}
\label{other}

It is often more useful to use the globally well-defined
gauge field $A$ defined in (\ref{globalA}) to describe the CS term.
A similar analysis as in (\ref{newCStr})--(\ref{newCStr2}) shows
that the new CS term (\ref{eqn:newCS}) can be rewritten as
\begin{eqnarray}
S_{\rm CS}^{\rm new}
&=&
C\Bigg(\int_{M_5}
\omega_5(A)
+\int_{N_5^{(+\infty)}}\frac{1}{10}\tr\left((h_+^{-1} dh_+)^5\right)
+\int_{M_4^{(+\infty)}}\alpha_4(d\wh h_+ \wh h_+^{-1}, A)
\nn\\
&&
-\int_{N_5^{(-\infty)}}\frac{1}{10}\tr\left((h_-^{-1} dh_-)^5\right)
-\int_{M_4^{(-\infty)}}\alpha_4(d\wh h_-\wh h_-^{-1}, A)
\Bigg)\ ,
\label{eqn:newCS2}
\end{eqnarray}
where $N_5^{(\pm\infty)}$ are 5 dimensional manifolds with
$\del N_5^{(\pm\infty)}=M_4^{(\pm\infty)}$ and $h_\pm$ are $U(N_f)$
valued function on $N_5^{(\pm\infty)}$ satisfying
$h_\pm|_{\del N_5^{(\pm\infty)}}=\wh h_\pm$.
The relation between the boundary values of the gauge field $A$ and the
external gauge fields $\wh A_\pm$ is given by (\ref{Abdry}).
The boundary terms in (\ref{eqn:newCS2}) can also be written in terms of
the external gauge fields as
\begin{eqnarray}
S_{\rm CS}^{\rm new}
&=&
C\Bigg(\int_{M_5}
\omega_5(A)
+\int_{N_5^{(+\infty)}}\frac{1}{10}\tr\left((h_+^{-1} dh_+)^5\right)
-\int_{M_4^{(+\infty)}}\alpha_4(d\wh h_+^{-1}\wh h_+,\wh A_+)
\nn\\
&&
-\int_{N_5^{(-\infty)}}\frac{1}{10}\tr\left((h_-^{-1} dh_-)^5\right)
+\int_{M_4^{(-\infty)}}\alpha_4(d\wh h_-^{-1}\wh h_-,\wh A_-)
\Bigg)\ ,
\label{eqn:newCS3}
\end{eqnarray}
where we have used (\ref{alpha4gauge}). This expression makes it clear
that we do not have to modify the CS term for $N_f=2$ and $\wh A_\pm=0$,
because the additional terms in (\ref{eqn:newCS3}) vanish in that case.

The expressions (\ref{eqn:newCS2}) and (\ref{eqn:newCS3})
can be written in a more compact notation as
\begin{eqnarray}
S_{\rm CS}^{\rm new}
&=&
C\left(\int_{M_5}
\omega_5(A)
+\int_{N_5}\frac{1}{10}\tr\left((h^{-1}d h)^5\right)
+\int_{\del M_5}\alpha_4(dh h^{-1}, A)
\right)\nn\\
&=&
C\left(\int_{M_5}
\omega_5(A)
+\int_{N_5}\frac{1}{10}\tr\left((h^{-1}dh)^5\right)
-\int_{\del M_5}\alpha_4(dh^{-1} h,\wh A)
\right)\, ,
\label{eqn:genCS2}
\end{eqnarray}
where $N_5$ is a 5 dimensional manifold with two connected components
$N_5=N_5^{(+\infty)}\cup (-N_5^{(-\infty)})$ satisfying
\begin{eqnarray}
\del N_5=\del M_5=M_4^{(+\infty)}\cup(-M_4^{(-\infty)})\ ,
\end{eqnarray}
and $h$ is a $U(N_f)$ valued function on $N_5$ with
$\wh h_\pm=h|_{M_4^{(\pm\infty)}}$.
The external gauge field $\wh A$ in (\ref{eqn:genCS2})
is defined on the boundary $\del M_5$ with the identification
$\wh A_\pm=\wh A|_{M_4^{(\pm\infty)}}$. The relation to the boundary
value (\ref{Abdry}) is written as
\begin{eqnarray}
A|_{\del M_5}=\wh A^{\,h} \,.
\label{Abdry2}
\end{eqnarray}

It is not difficult to show,
using (\ref{gauge5}), (\ref{alpha4gauge3}) and (\ref{eqn:v5-2}),
that this CS term is invariant
(up to a $2\pi\Z$ shift)
under the transformation (\ref{Ahtr}),
which can be written as
\begin{eqnarray}
A\ra A^g \ ,~~~h\ra gh\ ,~~\wh A\ra\wh A\ ,
\label{gaugetr2}
\end{eqnarray}
assuming that $g$ can be extended to $N_5$.

The transformation corresponding to the chiral symmetry
discussed around (\ref{chtr}) is given by
\begin{eqnarray}
 A\ra A^{g}\ ,~~~h\ra h\ ,~~~\wh A\ra\wh A^{\,\wh g} \,,
\label{chiraltr2}
\end{eqnarray}
with
\begin{eqnarray}
\wh g= (h^{-1} g h)|_{\del M_5}\ ,
\label{hgh}
\end{eqnarray}
where $\wh g_\pm\equiv \wh g|_{M_4^{(\pm\infty)}}$ corresponds
to the chiral symmetry. Combining this with the inverse of
(\ref{gaugetr2}), we find that
the chiral transformation is also induced by
\begin{eqnarray}
A\ra A\ ,~~~ h \ra  g^{-1}h\ ,~~~
\wh A\ra\wh A^{\,\wh g}\ .
\label{chiraltr2-2}
\end{eqnarray}

It is also straightforward to show that the CS term (\ref{eqn:genCS2})
transforms under the transformation (\ref{chiraltr2}) with (\ref{hgh})
as
\begin{eqnarray}
 S_{\rm CS}^{\rm new}\ra
 S_{\rm CS}^{\rm new}+C\left(
\int_{N_5}\frac{1}{10}\tr((\wh g d\wh g^{-1})^5)
+\int_{\del M_5}\alpha_4(d\wh g^{-1}\wh g,\wh A)
\right)\ ,
\label{chiraltr3}
\end{eqnarray}
up to $2\pi\Z$ shift,
where we have assumed that $\wh g$ can be extended to $N_5$.
If we consider an infinitesimal chiral transformation
with $\wh g\simeq 1-\wh\Lambda$, then (\ref{chiraltr3})
reduces to the formula for chiral anomaly (\ref{deltaSCS}).

There is another useful expression that generalizes (\ref{HMCS})
to the cases with boundary.
Note that $M_5\cup (-N_5)$ is a 5 dimensional manifold without
boundary.
Suppose there exists a 6 dimensional manifold $M_6$ 
with $\del M_6=M_5\cup (-N_5)$ and the gauge field $A$ can be extended
to $M_6$. Then, we have
\begin{eqnarray}
 \int_{M_6}\tr(F^3)
=\int_{M_5}\omega_5(A)-\int_{N_5}\omega_5(A)\ .
\label{F3}
\end{eqnarray}
Next, we extend the external gauge field $\wh A$ to $N_5$ by defining
$\wh A\equiv A^{h^{-1}}$ (on $N_5$), which reduces to (\ref{Abdry2})
at $\del N_5=\del M_5$. Then, using (\ref{gauge5}), we find
\begin{eqnarray}
 \int_{N_5}\omega_5(\wh A)= \int_{N_5}\left(\omega_5(A)
+\frac{1}{10}\tr\left((h^{-1}dh)^5\right)\right)
+\int_{\del N_5}\alpha_4(dhh^{-1},A)\ .
\label{omegawhA}
\end{eqnarray}
Comparing (\ref{F3}) and (\ref{omegawhA}) with (\ref{eqn:genCS2}), we obtain
a simple formula\footnote{A similar expression was
suggested in \cite{HM} as a quick remedy to recover
the chiral anomaly. Our derivation gives its precise meaning.
}
\begin{eqnarray}
 S_{\rm CS}^{\rm new}
=C\left(\int_{M_6}\tr(F^3)+\int_{N_5}\omega_5(\wh A)\right)\ .
\label{F3CS}
\end{eqnarray}

\subsection{Pion field}
\label{pion}

The relation between the $U(N_f)$ valued
pion field $U(x^\mu)$ in chiral Lagrangian and the 5
dimensional gauge field was proposed in
\cite{AM,Son:2003et,SS}:
\begin{eqnarray}
U(x^\mu)= {\rm P}
\exp\left(-\int_{-\infty}^{+\infty}dz A_z(x^\mu,z)\right)\ .
\end{eqnarray}
This formula should be modified as follows.

For the gauge field considered in section \ref{proposal1},
the correct expression is
\begin{eqnarray}
U(x^\mu)= {\rm P}
\exp\left(-\int_{0}^{+\infty}dz A_{+z}(x^\mu,z)\right)
h(x^\mu)|_{z=0}\, {\rm P}\exp\left(-\int_{-\infty}^0dz A_{-z}(x^\mu,z)\right)
\ .
\label{U1}
\end{eqnarray}
For the gauge field $A$ in (\ref{globalA}), this is equivalent to
\begin{eqnarray}
U(x^\mu)=
\wh h_+^{-1}(x^\mu)\,
{\rm P}
\exp\left(-\int_{-\infty}^{+\infty}dz A_{z}(x^\mu,z)\right)
\wh h_-(x^\mu)
\ .
\label{U2}
\end{eqnarray}
This expression is invariant under the gauge transformation
(\ref{Ahtr}).

On the other hand, (\ref{U1}) transforms under the gauge transformation
(\ref{gauge}) as
\begin{eqnarray}
U(x^\mu)\ra \wh g_+(x^\mu) U(x^\mu) \wh g_-(x^\mu)^{-1}\ ,
\label{chiraltr}
\end{eqnarray}
where $\wh g_\pm\equiv g_\pm|_{z\ra\pm\infty}$, which is nothing
but the chiral transformation of the pion field.
In terms of (\ref{U2}), (\ref{chiraltr}) can be easily seen
by the transformation (\ref{chiraltr2}) or (\ref{chiraltr2-2}).

\subsection{Equations of motion and current}
\label{eom-current}

For later use, let us write down the equations of motion
and currents with our new CS term. Since the additional terms in our new
CS term does not affect these equations, the results in this subsection
are not new. Nevertheless, it will be instructive to show them explicitly.
The action (\ref{S5dim}) is replaced with
\begin{eqnarray}
S_{\rm 5dim}=S_{\rm YM}+S_{\rm CS}^{\rm new}\ .
\label{S5dim2}
\end{eqnarray}
Here, we use the expression (\ref{eqn:genCS2}) for the CS
term $S_{\rm CS}^{\rm new}$.
Using (\ref{deltaomega}) and (\ref{deltaalpha}),
an infinitesimal variation of the action is computed as\footnote{
The variation with respect to $h$ can be absorbed in $\delta A$, using
the transformation (\ref{gaugetr2}).}
\begin{eqnarray}
\delta S= \int_{M_5}\tr\left(\delta A(-\kappa
D_A*\!F+3CF^2)\right)
+\int_{\del M_5}
\tr\left( \delta\wh A
\left(-\kappa\, \wh{*F}+C\left(\wh F\wh A+\wh A\wh F-\half \wh A^{\,3}
\right)
\right)
\right)\, , \nn \\
\label{eqn:deltaScurrent}
\end{eqnarray}
where $D_A$ is the covariant derivative defined in (\ref{covder}),
$\wh A$ is the external gauge field related to the boundary value
of the gauge field as (\ref{Abdry2}), and
\begin{eqnarray}
\wh F\equiv (h^{-1}F h)|_{\del M_5}\ ,~~~
\wh {*F}\equiv (h^{-1}*\!Fh)|_{\del M_5}\ ,~~~
\delta \wh A\equiv (h^{-1}\delta Ah)|_{\del M_5}\ .
\end{eqnarray}
Note here that $\wh{*F}$ is different from
the Hodge dual of $\wh F$ defined on $\del M_5$. Its explicit form with
(\ref{eqn:starF}) is
\begin{eqnarray}
 \wh{*F}=
\left(-\frac{k(z)}{3}(h^{-1} F^\mu_{~z}h)\epsilon_{\mu\nu\rho\sigma}
dx^\nu dx^\rho dx^\sigma\right)\Big|_{\del M_5}\ .
\end{eqnarray}

The first term in (\ref{eqn:deltaScurrent}) gives the equations
of motion
\begin{eqnarray}
-\kappa\, D_A*\!F+3CF^2=0 \, ,
\label{eqn:eom}
\end{eqnarray}
which is consistent with the boundary condition $\delta\wh A=0$.
The variation with respect to the external
gauge field $\wh A$ at the boundary in  (\ref{eqn:deltaScurrent}) gives 
the current 3-form :
\begin{eqnarray}
J_\pm\equiv \pm i\left(
-\kappa\, \wh{*F_\pm}+ C\left(\wh F_\pm\wh A_\pm+\wh A_\pm\wh F_\pm
-\half\wh A_\pm^{\,3}\right)
\right)
\ ,
\label{J2}
\end{eqnarray}
where $J_-$ and $J_+$ correspond to the currents of $U(N_f)_L$ and
$U(N_f)_R$, respectively
\cite{Hashimoto:2008zw,Rebhan:2008ur,Rebhan:2009vc}.
Then, it is straightforward to check, using the equations of motion
(\ref{eqn:eom}), it satisfies the (consistent)
anomaly equation:\footnote{
See \cite{Rebhan:2009vc} for a detailed discussion on
the currents and the anomaly equations in holographic QCD.
}
\begin{eqnarray}
D_{\wh A_\pm} J_\pm
=\pm \frac{N_c}{24\pi^2}d\left(\wh A_\pm d_\pm\wh A_\pm
+\half\wh A_\pm^{\,3}\right)\, . 
\end{eqnarray}

The baryon number current is defined as
\begin{eqnarray}
J_B=\frac{1}{N_c}\left(\tr J_++\tr J_-\right) \,,
\end{eqnarray}
and the baryon number (for $\wh A=0$) is
\begin{eqnarray}
n_B=\int_{S^3}J_B
=
\frac{i}{N_c}
\int_{S^3}\left[
\tr\left(
-\kappa\, \wh{*F}
\right)\right]^{z=+\infty}_{z=-\infty}
=\frac{1}{8\pi^2}\int_{S^3\times\sR}\tr(F^2)\ ,
\label{nB5}
\end{eqnarray}
where we have used the equations of motion (\ref{eqn:eom})
and Stokes' theorem in the last step,
reproducing the expression in (\ref{nB}).

\section{Application to baryons}
\label{Baryon_spec}

In this section, we analyze the effective action
for the collective coordinates of the soliton solution
corresponding to baryon. We show that the term (\ref{WZW}) needed to
obtain the correct constraint (\ref{constraint}) is reproduced
by using the CS term proposed in the previous section.
This statement was already shown in \cite{HM} using
(\ref{HMCS}) for the $n_B=1$ case.
As we have seen in section \ref{other} that our CS term reduces to
(\ref{HMCS}) when $\wh A=0$, we should recover their result.
In our derivation, we will not use an
explicit solution corresponding to a baryon so that it can be
generalized to the cases with $n_B>1$.

\subsection{Collective coordinates}

In this subsection, we work in the $A_0=0$ gauge. We assume there exists
a solution of the equations of motion (\ref{eqn:eom}) with non-zero
baryon number $n_B$, denoted as
\begin{eqnarray}
A^{\rm cl}=A_M^{\rm cl} dx^M \, ,
\end{eqnarray}
where ``${\rm cl}$'' refers to a classical solution and $M=1,2,3,z$ is
the spatial index. We also assume that this gauge field
is globally well-defined and regular everywhere in $M_5$.

Here, we consider the cases with $\wh A_\pm=0$.
Then, for a finite energy solution, the gauge field
approaches a pure gauge configuration near the boundary as
\begin{eqnarray}
A^{\rm cl}\ra h_\pm^{\rm cl}d h_\pm^{{\rm cl}-1}\ ,~~~(z\ra\pm\infty)\ .
\end{eqnarray}
Because of the condition $A_0=0$, 
$h_\pm^{\rm cl}$ are time independent.
Without
loss of generality, we can assume $h_-^{\rm cl}|_{z\ra\infty}=1$
and $h_+^{\rm cl}|_{z\ra+\infty}\equiv h_0(\vec x)$,
where $h_0$ is a $U(N_f)$ valued function
on the $S^3$ parametrized by $\vec x=(x^1,x^2,x^3)$
satisfying
\begin{eqnarray}
n_B=\frac{1}{24\pi^2}\int_{S^3}\tr\left((h_0^{-1}dh_0)^3\right)\ .
\end{eqnarray}

Following \cite{HSSY}, we consider a gauge configuration
\begin{eqnarray}
A_M=VA_M^{\rm cl}V^{-1}+V\del_M V^{-1}
\label{eqn:A0_gaugecon}
\end{eqnarray}
with a globally well-defined $SU(N_f)$ valued function $V$.\footnote{
One could consider $V$ to be a $U(N_f)$ valued function. However,
we only consider the configurations of $V$ that do not wind around
a non-trivial 1-cycle of $U(N_f)$ along the time direction
in the following (see the footnote in p.12 for a related issue)
and, at least for such configurations, it is possible to show that 
the diagonal $U(1)$ part of the $U(N_f)$ does not contribute to the
effective action studied in section \ref{effaction} and we can restrict
$V$ to be an $SU(N_f)$ valued function.
}
The idea is as follows. If $V$ is time independent, it can be regarded
as the collective coordinates (coordinates of the instanton moduli
space) corresponding to the global gauge rotation,
since $A_M$ is again a classical solution with the same energy.
A standard procedure of the moduli space quantization
method\footnote{See, e.g., \cite{Harvey:1996ur} for a review of this
method explained for the magnetic monopoles.}
is to promote the collective coordinates to be time dependent variables
and reduce the system to a quantum mechanics of these variables.
To this end, one should also make a compensating gauge transformation so
that the gauge configuration satisfies the Gauss law equation, which is
the equation of motion for $A_0$:
\begin{eqnarray}
dt\wedge\left(-\kappa\, D_A*\!F+3CF^2\right)=0 \ . 
\label{eqn:GL1}
\end{eqnarray}
$V$ in (\ref{eqn:A0_gaugecon}) contains both the collective coordinates
and the compensating gauge transformation, and it can depend on
the 5 dimensional space-time coordinates.
We assume that the initial value of $V$ is $1$ and hence its value at a
fix time is connected to $V=1$ by a continuous deformation.

With this choice of the gauge configuration, the asymptotic value of the
gauge field is
\begin{eqnarray}
A_M\ra Vh_\pm^{\rm cl}\del_M(Vh_\pm^{\rm cl})^{-1}\ ,~~~
 (z\ra\pm\infty)  \ .
\end{eqnarray}
The electric fields $F_{0i}$ ($i=1,2,3$) are assumed to vanish at the
boundaries $z\ra\pm\infty$. Then,
$F_{0i}|_{z\ra\pm\infty}= \partial_0 A_i|_{z\ra\pm\infty}=0$
implies that the asymptotic values of $A_i$ should be time
independent, and therefore, since the initial value of $V$ is assumed to
be $1$, one has
\begin{eqnarray}
A_i\ra
h_\pm^{\rm cl}\del_i h_\pm^{\rm cl\,-1}
 \ ,~~~  (z\ra\pm\infty)
\label{eqn:asymA}
\end{eqnarray}
for all time.
This implies that $V$ has the following asymptotic values
\begin{eqnarray}
V|_{z\ra-\infty}=a_-(t)\ ,~~~
V|_{z\ra+\infty}=h_0(\vec x)a_+(t)h_0^{-1}(\vec x)\ ,
\label{eqn:Vbdry}
\end{eqnarray}
with $a_\pm(t)$ being $SU(N_f)$ valued functions that depend only
on time.

With the asymptotic expression of the gauge field in (\ref{eqn:asymA}),
$\wh h_\pm$ in (\ref{Abdry}) can be chosen as
\begin{eqnarray}
\wh h_-=1\, ,~~\wh h_+=h_0(\vec x) \,,
\end{eqnarray}
and the CS term (\ref{eqn:newCS3}) is simply
\begin{eqnarray}
S_{\rm CS}^{\rm new}=C\int_{M_5}\omega_5(A)\ .
\label{eqn:CS-0}
\end{eqnarray}
Therefore, the naive CS term is actually the correct one in this gauge choice.

Let us now consider the Gauss law equation (\ref{eqn:GL1}).
With the expression (\ref{eqn:A0_gaugecon}), one can easily show that
$F_{MN}=VF_{MN}^{\rm cl}V^{-1}$ and
\begin{eqnarray}
F_{0M}=\dot A_M=V\left(F_{0M}^{\rm cl}-D_M^{\rm cl}\Phi
\right)V^{-1}\, ,
\end{eqnarray}
where dot denotes the time derivative, and we have defined
$\Phi\equiv V^{-1}\dot V$ and $D_M^{\rm cl}\Phi\equiv
\del_M\Phi+[A_M^{\rm cl},\Phi]$. Using these relations and the fact that
$A_M^{\rm cl}$ is a classical solution, (\ref{eqn:GL1}) becomes
\begin{eqnarray}
dt\wedge (D_A^{\rm cl}*\!(D_A^{\rm cl}\Phi\, dt))=0\ ,
\label{eqn:GL2}
\end{eqnarray}
where the covariant derivative acting on $\Phi$ is
$D_A^{\rm cl}\Phi\equiv D_M^{\rm cl}\Phi\, dx^M$.
In components, (\ref{eqn:GL2}) is given by
\begin{eqnarray}
D_M^{\rm cl}(\sqrt{-g}\, g^{MN}g^{00} D_N^{\rm cl}\Phi)=0\ .
\end{eqnarray}
For the background with the metric (\ref{eqn:metric}), this is written
explicitly as
\begin{eqnarray}
\delta^{ij} D_i^{\rm cl} D_j^{\rm cl}\Phi
+\wt k(z)^{-1} D_z^{\rm cl}(k(z) D_z^{\rm cl}\Phi)=0\ ,
\end{eqnarray}
where $i,j=1,2,3$.

With the expression (\ref{eqn:Vbdry}),
$\Phi$ has the following asymptotic values
\begin{eqnarray}
 \Phi|_{z\ra-\infty}=a_-(t)^{-1}\dot a_-(t)\ ,~~~
 \Phi|_{z\ra+\infty}=
h_0(\vec x)a_+(t)^{-1}\dot a_+(t)h_0^{-1}(\vec x)\ .
\label{eqn:bdryPhi}
\end{eqnarray}
Therefore, $\Phi$ is determined as the solution of
the Gauss law equation (\ref{eqn:GL2}) with
the boundary condition (\ref{eqn:bdryPhi}).

\subsection{Effective action}
\label{effaction}

To obtain the effective action for $a_\pm(t)$,
it turns out to be more convenient to
make a gauge transformation (\ref{Ahtr})
using $g=V^{-1}$. Then,
the configuration in (\ref{eqn:A0_gaugecon}) is mapped to
\begin{eqnarray}
A_0=V^{-1}\dot V\equiv \Phi\ ,~~~A_M=A_M^{\rm cl}\ ,
\label{APhi}
\end{eqnarray}
and $\wh h_\pm$ in (\ref{Abdry}) is given by
\begin{eqnarray}
\wh h_-=a_-(t)^{-1}\ ,~~\wh h_+=h_0(\vec x)a_+(t)^{-1} \ .
\label{handa}
\end{eqnarray}
Then, the CS term (\ref{eqn:newCS3}) is
\begin{eqnarray}
S_{\rm CS}^{\rm new}
=C\left(\int_{M_5}\omega_5(A)+
\int_{N_5^{(+\infty)}}\frac{1}{10}\tr\left((a_+h_0^{-1} d(h_0a_+^{-1}))^5 \right)
\right)\ .
\label{eqn:CS-Phi}
\end{eqnarray}
Here, $N_5^{(+\infty)}$ is assumed to be
$N_5^{(+\infty)}\simeq D\times S^3$, and $h_0$ and $a_+$
are extended to be functions on it. We can choose
$h_0$ and $a_+$ to be constant along the $D$ and $S^3$
directions, respectively.
Using the relation (\ref{eqn:v5-2}), one can show that
(\ref{eqn:CS-Phi}) is equivalent to
\begin{eqnarray}
S_{\rm CS}^{\rm new}
=C\left(\int_{M_5}\omega_5(A)
-\frac{1}{2}\int_{M_4^{(+\infty)}}
dt \tr\left(a_+^{-1}\dot a_+ \, (h_0^{-1} dh_0)^3 \right)
\right)\,. 
\label{eqn:action_phigauge}
\end{eqnarray}
Although it is a bit more tedious, it is also possible to derive
(\ref{eqn:action_phigauge}) directly from (\ref{eqn:CS-0}) by using
(\ref{gauge5}) with $g=V^{-1}$.\footnote{
The integral of $\tr((V^{-1} dV)^5)$ over $M_5$ can be evaluated by
using (\ref{trick}).
}

The first term on the right hand side of (\ref{eqn:action_phigauge})
can be evaluated as follows. The relation (\ref{deltaomega}) with
$\delta A=\Phi \, dt$ implies 
\begin{eqnarray}
\omega_5(A)=\omega_5(A^{\rm cl})
+3\tr(\Phi dt \, (F^{\rm cl})^2)+d\beta_4(\Phi dt,A^{\rm cl})\ ,
\end{eqnarray}
where $\beta_4$ is defined in (\ref{beta4}).
The contribution from the collective coordinates to the CS 5-form is 
\begin{eqnarray}
&& \int_{M_5}\omega_5(A)-\int_{M_5}\omega_5(A^{\rm cl})
\nn\\
&=& \int_{M_5}3\tr(\Phi dt \,(F^{\rm cl})^2)
+\int_{M_4^{(+\infty)}}\beta_4(\Phi dt,A^{\rm cl})|_{z=+\infty}
-\int_{M_4^{(-\infty)}}\beta_4(\Phi dt,A^{\rm cl})|_{z=-\infty}
\nn\\
&=& \int_{M_5}3\tr(\Phi dt \, (F^{\rm cl})^2)
+\half\int_{M_4^{(+\infty)}} dt\tr\left(a_+^{-1}\dot a_+ 
(h_0^{-1}dh_0)^3\right)\ .
\end{eqnarray}
Substituting this back to (\ref{eqn:action_phigauge}), one obtains
\begin{eqnarray}
S_{\rm CS}^{\rm new}=
\int_{M_5}\omega_5(A^{\rm cl})+
3C\int_{M_5}dt \tr\left(\Phi  (F^{\rm cl})^2\right)\ .
\end{eqnarray}

The field strength for the gauge field (\ref{APhi}) is
\begin{eqnarray}
F=F^{\rm cl}+D_A^{\rm cl}\Phi dt\, ,
\end{eqnarray}
and the YM part is given as
\begin{eqnarray}
S_{\rm YM}
&=&
S_{\rm YM}(A^{\rm cl})
-\frac{\kappa}{2}\int_{M_5}\tr\left(
D_A^{\rm cl}\Phi dt\wedge *(D_A^{\rm cl}\Phi dt)
\right)
\nn\\
&&
-\kappa
\int_{M_5}dt\,\tr\left(\Phi\,D_A^{\rm cl} *
F^{\rm cl}\right)
-\kappa\int_{\del M_5}dt\tr\left(\Phi *\!F^{\rm cl}\right)\ .
\end{eqnarray}
Using the fact that $A^{\rm cl}$ satisfies the equations of motion
(\ref{eqn:eom}),
the total action (\ref{S5dim2}) becomes
\begin{eqnarray}
S_{\rm 5dim}= S_{\rm 5dim}(A^{\rm cl})+S_1+S_2\ ,
\label{SQM}
\end{eqnarray}
where $S_{\rm 5dim}(A^{\rm cl})$ is the action evaluated
with $A=A^{\rm cl}$,
$S_1$ and $S_2$ are the terms including $\Phi$:
\begin{eqnarray}
S_1&=&-\kappa\int_{\del M_5}dt\tr\left(\Phi *\!F^{\rm cl}\right)
\ ,
\label{S1}\\
S_2&=&-\frac{\kappa}{2}\int_{M_5}\tr\left(
D_A^{\rm cl}\Phi dt\wedge *(D_A^{\rm cl}\Phi dt)
\right)\ .
\label{S2}
\end{eqnarray}
Using the Gauss law equation (\ref{eqn:GL2}),
$S_2$ can also be written as
\begin{eqnarray}
 S_2=-\frac{\kappa}{2}\int_{\del M_5}
 dt\tr\left(\Phi*\!(D_A^{\rm cl}\Phi dt)\right)\ .
\label{S2-2}
\end{eqnarray}
For the background with the metric (\ref{eqn:metric}), (\ref{S1})
and (\ref{S2-2}) can be written as
\begin{eqnarray}
S_1&=&2\kappa\int d^4x\left[k(z)\tr\left(\Phi F_{0z}^{\rm cl}
\right)\right]^{z\ra+\infty}_{z\ra-\infty}
\ ,\\
S_2&=&\kappa\int d^4x\left[k(z)\tr\left(
\Phi D_z^{\rm cl}\Phi
\right)\right]^{z\ra+\infty}_{z\ra-\infty}\ .
\end{eqnarray}

Substituting the asymptotic expressions of $\Phi$
(\ref{eqn:bdryPhi}) into (\ref{S1}), one obtains
\begin{eqnarray}
S_1
=-i\int dt \tr\left(
a_+^{-1}\dot a_+ n_+^{\rm cl}
+a_-^{-1}\dot a_- n_-^{\rm cl}
\right)\ ,
\label{S1-2}
\end{eqnarray}
with $n_\pm^{\rm cl}$ defined by
\begin{eqnarray}
 n_\pm^{\rm cl}\equiv \int_{S^3} J_\pm^{\rm cl}
= \mp i\kappa\int_{S^3} \wh{*F^{\rm cl}}|_{z\ra\pm\infty}\ ,
\label{npm}
\end{eqnarray}
where $J_\pm^{\rm cl}$ are the classical current 3-forms
given by (\ref{J2}) with $A=A^{\rm cl}$ and $\wh A_\pm=0$.
The classical quark number matrix is defined as
$n_Q^{\rm cl}\equiv n_+^{\rm cl}+n_-^{\rm cl}$. Its diagonal elements
are interpreted as the number of up quarks, down quarks, strange quarks,
etc., carried by the classical solution
and the trace is the total quark number:
\begin{eqnarray}
 \tr n_Q^{\rm cl}= N_c n_B\ .
\end{eqnarray}

\subsection{Relation to Skyrmions}
\label{skyrme}

The action of the Skyrme model is written in terms of the pion field
$U(x^\mu)$ discussed in section \ref{pion}. The classical solution
corresponding to the baryon carries non-zero winding number as
an element of $\pi_3(U(N_f))\simeq\Z$.
In the standard approach for $N_f=3$, the ansatz for the field
configuration is
\begin{eqnarray}
 U(x^\mu)=a(t)U^{\rm cl}(\vec x) a(t)^{-1}\ ,
\label{Skyrmion}
\end{eqnarray}
where $U^{\rm cl}(\vec x)\in SU(3)$ is a classical solution
representing a baryon and $a(t)\in SU(3)$ is the
collective coordinates corresponding to the $SU(3)$ rotation.
The classical solution is assumed to be of the form
\begin{eqnarray}
 U^{\rm cl}(\vec x)=
\left(
\begin{array}{cc}
U_0(\vec x) & \\
&1
\end{array}
\right)\ ,
\label{U0}
\end{eqnarray}
where $U_0(\vec x)$ is the Skyrmion solution for $N_f=2$. The form of
the solution (\ref{U0}) is natural in the sense that exciting the
components of the mesons with a strange quark costs more energy than
those with only up and down quarks, when we include the mass term
to the Lagrangian.

The pion field (\ref{U2}) for
our gauge configuration (\ref{APhi}) is given by
\begin{eqnarray}
U(x^\mu)=
a_+(t) h_0^{-1}(\vec x)
{\rm P}
\exp\left(-\int_{-\infty}^{+\infty}dz A^{\rm cl}_{z}(x^\mu,z)\right)
a_-(t)^{-1}
\ ,
\end{eqnarray}
and it corresponds to the above ansatz (\ref{Skyrmion})
with the identification $a_+(t)=a_-(t)=a(t)$ and
\begin{eqnarray}
U^{\rm cl}(x^\mu)=
 h_0^{-1}(\vec x)
{\rm P}
\exp\left(-\int_{-\infty}^{+\infty}dz A^{\rm cl}_{z}(x^\mu,z)\right)
\ .
\label{Ucl}
\end{eqnarray}
Note that, in the infinite volume limit, the pion field is supposed to
approach its vacuum value at spatial infinity, {\it i.e.}
 $U(x^\mu)|_{|\vec x|\ra\infty}=1$.
Since the modes with $a_+\ne a_-$ change the vacuum configuration, they
are unphysical in the infinite volume limit. For this reason,
we impose $a_+=a_-$ hereafter.

Motivated by the ansatz (\ref{U0}), we consider embedding a classical
solution for $N_f=2$ into the $U(3)$ gauge field to obtain $A^{\rm cl}$
for $N_f=3$, as it was done in \cite{HM}. Decomposing
the $U(2)$ gauge field into the $SU(2)$ part and $U(1)$ part as
\begin{eqnarray}
 A^{U(2)}=A^{SU(2)}+A^{U(1)}\ ,
\end{eqnarray}
the equations of motion (\ref{eqn:eom}) for $N_f=2$ can be written as
\begin{eqnarray}
&& -\kappa\, D_A*\!F^{SU(2)}+6 C F^{U(1)}F^{SU(2)}=0\ ,
\\
&& -\kappa\, d*\!F^{U(1)}+3 C\left((F^{U(1)})^2+(F^{SU(2)})^2\right)=0
\ .
\label{eomU1}
\end{eqnarray}
These equations can be consistently truncated by restricting
$F_{0M}^{SU(2)}=0$ and $F_{MN}^{U(1)}=0$ for $M,N=1,2,3,z$.
In this case, only the $U(1)$ part of the gauge field contributes
in (\ref{npm}) and the classical quark number matrix $n_Q^{\rm cl}$
for $N_f=2$ is proportional to the unit matrix. When the solution for
$N_f=2$ is embedded into the $U(3)$ gauge field, $n_Q^{\rm cl}$ is of
the form
\begin{eqnarray}
 n_Q^{\rm cl}
=\frac{N_c n_B}{2}
\left(
\begin{array}{ccc}
1 & & \\
 &1 & \\
 & & 0\\
\end{array}
\right)\ ,
\label{nQ}
\end{eqnarray}
which means that,
before quantization of the collective modes $a(t)$,
the classical configuration represents a state with
no strangeness and equal number of up and down quarks.

Imposing $a_+=a_-\equiv a\in SU(3)$, (\ref{S1-2}) becomes
\begin{eqnarray}
S_1
=-i\int dt \tr\left(
a^{-1}\dot a\, n_Q^{\rm cl}
\right)
=-i\frac{N_cn_B}{\sqrt{3}}\int dt \tr\left(t_8
a^{-1}\dot a\,
\right)
\ ,
\label{S1-3}
\end{eqnarray}
which precisely agrees with (\ref{WZW}).
Note that $\tr(t_8a^{-1}\dot a)$ does not appear in $S_2$.
To see this, let us assume that $a$ is of the form $a=e^{it_8\theta(t)}$.
For this, since $t_8$ commutes with $A^{\rm cl}$ and $h_0$,
$\Phi=a^{-1}\dot a=it_8\dot\theta$ solves the equations (\ref{eqn:GL2})
and (\ref{eqn:bdryPhi}). Then, it is clear that $S_2$ vanishes.
Because $\dot\theta$ appears only in (\ref{S1-3}), the momentum
conjugate to $\theta$ is
\begin{eqnarray}
 P_\theta=\frac{N_cn_B}{2\sqrt{3}}\ ,
\end{eqnarray}
and hence the correct baryon constraint (\ref{constraint})
is recovered.

\section{Conclusion and outlook}
\label{Conclusion}

In this paper, we re-examined a puzzle concerned with the CS term
in the 5 dimensional meson effective theory of holographic QCD.
We proposed a modified CS term and demonstrated that the new
action successfully reproduces the required baryon constraint as well as
the chiral anomaly.

Although we obtained a CS term that can be used for the topologically
non-trivial gauge configurations corresponding to baryons,
our construction is not completely general.
For example, the expression (\ref{eqn:genCS2}) is applicable
only when $N_5$ and $h$ can be constructed and the gauge field can be
treated as globally well-defined 1-form field on $M_5$.
For the expression (\ref{F3CS}), we have to assume the existence of
$M_6$ and $N_5$ as well as an extension of the gauge fields
to these spaces. (See the footnote in p.12 for further comments.) 
It would be interesting to investigate an expression of the CS
term that works for more generic situation, as it was done
in \cite{Dijkgraaf:1989pz} for the 3 dimensional CS term.

The main motivation for the present work is to solve
a puzzle concerned with baryons in holographic QCD with $N_f=3$
and make it applicable to the physics of baryons including strange
quarks. In order to be more realistic, it would be important
to include the mass of the strange quark. There are already some works
along this direction. (See, e.g.,
\cite{HM,Hashimoto:2009st,Ishii:2010ib,
Aoki:2012th,Matsumoto:2016lxa,Suganuma:2016lmp})
We hope our work removes possible concerns on the validity
of the formulation and provides some new insight into application
of holographic QCD to hyperons.

\section*{Acknowledgement}

We thank K. Hashimoto and H. Hata for discussion, and
Y. Kikuchi for the collaboration in the early stage of the project.
P.H.C. Lau acknowledges support as an International Research Fellow of 
the Japan Society for the Promotion of Science (JSPS).
The work of S.S was supported by JSPS KAKENHI
(Grant-in-Aid for Scientific Research (C))
Grant Number JP16K05324.

\appendix

\section{Notations and useful formulae}
\label{notations}

\subsection{Gauge field, covariant derivative, etc.}
In our convention, the gauge field $A$
and its field strength $F=dA+A^2$ are
anti-Hermitian 1-form and 2-form, respectively.
The gauge transformation is
\begin{eqnarray}
 A\ra A^g\equiv gAg^{-1}+gdg^{-1}=g(A+dg^{-1}g)g^{-1}
\ ,~~~F\ra F^g\equiv gFg^{-1}\ .
\end{eqnarray}
For a general (matrix valued) $n$-form $\alpha_n$,
we define $D_A\alpha_n$ as
\begin{eqnarray}
 D_A\alpha_n\equiv d\alpha_n+A\alpha_n-(-1)^n\alpha_n A\ .
\label{covder}
\end{eqnarray}
It satisfies Leibniz rule
\begin{eqnarray}
 D_A(\alpha_n\beta_m)= (D_A\alpha_n)\beta_m+(-1)^n \alpha_n D_A\beta_m\ .
\end{eqnarray}
One can show
\begin{eqnarray}
 D_A F= dF+AF-FA =0\ .
\end{eqnarray}
Note that $d$ and $D_A$ are the same in the trace:
\begin{eqnarray}
d\tr \alpha_n =\tr d\alpha_n=\tr (D_A\alpha_n)\ .
\end{eqnarray}

The infinitesimal variation of the field strength is
\begin{eqnarray}
\delta F=d\delta A+\delta A A+A\delta A+\delta A^2
= D_A\delta A
+\cO(\delta A^2)\ . 
\end{eqnarray}

The infinitesimal gauge transformation with
$g=e^{-\Lambda}$ is
\begin{eqnarray}
\delta_\Lambda A
\equiv (A^g-A)|_{\cO(\Lambda)}
=d\Lambda+[A,\Lambda]= D_A\Lambda\ .
\end{eqnarray}

The following trivial relations that follows from
$\tr((\mbox{odd form})^{2n})=0$
are sometimes useful:
\begin{eqnarray}
 \tr(A^2)=\tr(A^4)=0\ ,~~~\tr(AFAF)=0\ .
\end{eqnarray}

\subsection{CS 3-form}

The CS 3-form is defined as
\begin{eqnarray}
 \omega_3(A)
\equiv\tr\left(AF-\frac{1}{3} A^3
\right)
=\tr\left(AdA+\frac{2}{3} A^3
\right)\ ,
\end{eqnarray}
which satisfies
\begin{eqnarray}
 d\omega_3(A)=\tr(F^2)\ .
\end{eqnarray}

The gauge transformation is
\begin{eqnarray}
\omega_3(A^g)=\omega_3(A)-\frac{1}{3}\tr((gdg^{-1})^3)
-d \tr(dg^{-1} gA)\ .
\end{eqnarray}
The infinitesimal gauge transformation with
$g=e^{-\Lambda}$ and
$\delta_\Lambda A=D_A\Lambda$ is
\begin{eqnarray}
 \delta_\Lambda\omega_3(A)=d\tr\left(\Lambda dA\right)
+\cO(\Lambda^2)\ .
\end{eqnarray}

The infinitesimal variation is
\begin{eqnarray}
 \delta\omega_3(A)=2\tr(\delta A F)
+d\tr(\delta A A)+\cO(\delta A^2)\ .
\end{eqnarray}

\subsection{CS 5-form}
\label{CS5form}

The definition of the CS 5-form is
\begin{eqnarray}
\omega_5(A)\equiv\tr\left(
AF^2-\half A^3F+\frac{1}{10}A^5
\right)
=\tr\left(
AdAdA
+\frac{3}{2}A^3dA
+\frac{3}{5}A^5
\right) \ ,
\end{eqnarray}
which satisfies
\begin{eqnarray}
 d\omega_5(A)=\tr(F^3)\ .
\end{eqnarray}

The gauge transformation is
\begin{eqnarray}
 \omega_5(A^g)=\omega_5(A)+\frac{1}{10}\tr((gdg^{-1})^5)
+d\alpha_4(dg^{-1}g,A)\ ,
\label{gauge5}
\end{eqnarray}
where
\begin{eqnarray}
 \alpha_4(V,A)&=&
-\half\tr\left(V(AdA+dAA+A^3)-\half VAVA-V^3A\right)
\nn\\
&=&
\half\tr\left(V(A^3-AF-FA)+\half VAVA+V^3A\right)
\ .
\end{eqnarray}
This $\alpha_4(V,A)$ satisfies the following relations:
\begin{eqnarray}
 \alpha_4(V,\pm V)=0
\end{eqnarray}
for any one form $V$,
\begin{eqnarray}
 \alpha_4(dgg^{-1},A^g)=-\alpha_4(dg^{-1}g,A)\ ,
\label{alpha4gauge}
\end{eqnarray}
and
\begin{eqnarray}
\alpha_4(d(gh)(gh)^{-1},A^g)
&=& \alpha_4(g(H-G)g^{-1},A^g)\nn\\
&=&
\alpha_4(H,A)-\alpha_4(G,A)
-\half\tr\left(G^3H+GH^3-\half GHGH\right)
\nn\\
&&
+\half d\tr\left(
(H-G)(AG-GA)
\right)
\ ,
\label{alpha4gauge2}
\end{eqnarray}
where $G=dg^{-1}g$ and $H=dh h^{-1}$.
Using (\ref{alpha4gauge}) and (\ref{alpha4gauge2}), one can also show
\begin{eqnarray}
\alpha_4(d(gh)^{-1}(gh),A)
&=&\alpha_4(dh^{-1}h,A)+\alpha_4(G,A^h)
+\half\tr\left(G^3H+GH^3-\half GHGH
\right)
\nn\\
&&-\half d \tr\left((H-G)(A^hG-GA^h))\right)\ ,
\label{alpha4gauge3}
\end{eqnarray}
where $G=dg^{-1}g$ and $H=dh h^{-1}$.

The infinitesimal variation is
\begin{eqnarray}
\delta \omega_5(A)=3\tr(\delta A F^2)+d\beta_4(\delta A,A)+
\cO(\delta A^2)\ ,
\label{deltaomega}
\end{eqnarray}
where
\begin{eqnarray}
\beta_4(\delta A,A) \equiv\tr\left(
\delta A\left(FA+AF-\half A^3\right)
\right)\ .
\label{beta4}
\end{eqnarray}

The infinitesimal gauge transformation with $g=e^{-\Lambda}$ and
$\delta_\Lambda A= D_A\Lambda$ is
\begin{eqnarray}
\delta_\Lambda\omega_5(A)|_{\cO(\Lambda)}=
d\alpha_4(d\Lambda,A)|_{\cO(\Lambda)}=d\omega^1_4(\Lambda,A)
=d\left(3\tr(\Lambda F^2)+\beta_4(D_A\Lambda,A)\right)\ ,
\end{eqnarray}
where
\begin{eqnarray}
\omega^1_4(\Lambda,A)
&\equiv&\tr\left(\Lambda\, d\left(AdA+\half A^3\right)\right)\nn\\
&=&\half\tr\left(\Lambda\, \left(2F^2-FA^2-AFA-A^2F+A^4\right)\right)
\ .
\label{omega14}
\end{eqnarray}

The infinitesimal variation of $\alpha_4(A)$ is
\begin{eqnarray}
\delta \alpha_4(V,A)&\equiv&
\alpha_4(V,A+\delta A)-\alpha_4(V,A)
\nn\\
&=&\half\tr\left(
\delta A(2FV+2VF-(A+V)^3+A^3)
\right)
-\half d\tr\left(\delta A\, [V,A]\right)
+\cO(\delta A^2)\ .
\nn\\
\label{deltaalpha}
\end{eqnarray}

\subsection{WZW}

When $U=gh$, where $g$ and $h$ are $U(N_f)$ valued functions, we have
\begin{eqnarray}
\tr((U^{-1}dU)^3)=-\tr(G^3)+\tr(H^3)
+3\,d\tr(GH)\ ,
\end{eqnarray}
and
\begin{eqnarray}
 \tr((U^{-1}dU)^5)=-\tr(G^5)+\tr(H^5)
+5\,d\tr\left(G^3H+GH^3-\half GHGH
\right)\ ,
\label{eqn:v5-2}
\end{eqnarray}
where $G=dg^{-1}g$, $H=dhh^{-1}$.
This formula can also be shown from (\ref{gauge5})
by setting $U^{-1}dU=A^{h^{-1}}$ with $A=g^{-1}dg=-G$.

When $U=gfh$, where $g$, $f$ and $h$ are $U(N_f)$ valued functions,
we have
\begin{eqnarray}
\tr((U^{-1}dU)^5)
&=&
-\tr(G^5)+\tr(F^5)+\tr(H^5)
\nn\\
&&+5\,d\tr\Big(f^{-1}(G-F)^3fH+f^{-1}(G-F)fH^3
\nn\\
&&-\half(f^{-1}(G-F)fH)^2 +G^3F+GF^3-\half GFGF
\Big)\nn\\
&=&
-\tr(G^5)-\tr(\wh F^5)+\tr(H^5)
\nn\\
&&+5\,d\tr\Big(G^3f(H-\wh F)f^{-1}
+Gf(H-\wh F)^3f^{-1}
\nn\\
&&-\half(Gf(H-\wh F)f^{-1})^2 +\wh F^3H+\wh F H^3-\half \wh FH\wh FH
\Big)\ ,
\label{v5}
\end{eqnarray}
where $G=dg^{-1}g$, $F=df f^{-1}$, $\wh F=df^{-1}f$ and $H=dhh^{-1}$.

An important property is that when $M_5$ is a 5 dimensional closed
manifold, the integral
\begin{eqnarray}
 \frac{C}{10}\int_{M_5}\tr((U^{-1}dU)^5)
\end{eqnarray}
takes values in $2\pi\Z$
and its contribution in the action can be dropped.
When $M_5$ has a boundary, a useful trick to evaluate this integral
is to find $N_5$ such that $\del N_5=\del M_5$, {\it i.e.}
$M_5\cup (-N_5)$ is a closed manifold, and 
extend $U$ to be a $U(N_f)$-valued function on $M_5\cup (-N_5)$.
If such $N_5$ and $U$ exist, $M_5$ can be replaced
with $N_5$ by using
\begin{eqnarray}
 \frac{C}{10}\int_{M_5}\tr((U^{-1}dU)^5)
= \frac{C}{10}\int_{N_5}\tr((U^{-1}dU)^5)\ ,~~~(\mod 2\pi\Z)\ .
\label{trick}
\end{eqnarray}


\begin{thebibliography}{99}
	
\bibitem{Mald}
  J.~M.~Maldacena,
  ``The Large N limit of superconformal field theories and supergravity,''
  Int.\ J.\ Theor.\ Phys.\  {\bf 38} (1999) 1113
   [Adv.\ Theor.\ Math.\ Phys.\  {\bf 2} (1998) 231]
  [hep-th/9711200].

\bibitem{GKP}
  S.~S.~Gubser, I.~R.~Klebanov and A.~M.~Polyakov,
  ``Gauge theory correlators from noncritical string theory,''
  Phys.\ Lett.\ B {\bf 428} (1998) 105
  [hep-th/9802109].

\bibitem{Witt}
  E.~Witten,
  ``Anti-de Sitter space and holography,''
  Adv.\ Theor.\ Math.\ Phys.\  {\bf 2} (1998) 253
  [hep-th/9802150].

\bibitem{Guijosa:2016upo}
  A.~G\'uijosa,
  ``QCD, with strings attached,''
  Int.\ J.\ Mod.\ Phys.\ E {\bf 25} (2016) no.10,  1630006
  [arXiv:1611.07472 [hep-th]].

\bibitem{SS}
  T.~Sakai and S.~Sugimoto,
  ``Low energy hadron physics in holographic QCD,''
  Prog.\ Theor.\ Phys.\  {\bf 113} (2005) 843
  [hep-th/0412141].

\bibitem{Rebhan:2014rxa}
  A.~Rebhan,
  ``The Witten-Sakai-Sugimoto model: A brief review and some recent results,''
  EPJ Web Conf.\  {\bf 95} (2015) 02005
  [arXiv:1410.8858 [hep-th]].

\bibitem{Son:2003et}
  D.~T.~Son and M.~A.~Stephanov,
  ``QCD and dimensional deconstruction,''
  Phys.\ Rev.\ D {\bf 69} (2004) 065020
  [hep-ph/0304182].


%
%
%
%


\bibitem{DH}
  S.~K.~Domokos and J.~A.~Harvey,
  ``Baryon number-induced Chern-Simons couplings of vector and
	axial-vector mesons in holographic QCD,'' 
  Phys.\ Rev.\ Lett.\  {\bf 99} (2007) 141602
  [arXiv:0704.1604 [hep-ph]].

\bibitem{Pomarol:2008aa}
  A.~Pomarol and A.~Wulzer,
  ``Baryon Physics in Holographic QCD,''
  Nucl.\ Phys.\ B {\bf 809} (2009) 347
  [arXiv:0807.0316 [hep-ph]].

\bibitem{Domokos:2009cq}
  S.~K.~Domokos, H.~R.~Grigoryan and J.~A.~Harvey,
  ``Photoproduction through Chern-Simons Term Induced Interactions in
	Holographic QCD,'' 
  Phys.\ Rev.\ D {\bf 80} (2009) 115018
  [arXiv:0905.1949 [hep-ph]].

\bibitem{quarkmass}
  O.~Aharony and D.~Kutasov,
  ``Holographic Duals of Long Open Strings,''
  Phys.\ Rev.\ D {\bf 78} (2008) 026005
  [arXiv:0803.3547 [hep-th]].
\\
  K.~Hashimoto, T.~Hirayama, F.~L.~Lin and H.~U.~Yee,
  ``Quark Mass Deformation of Holographic Massless QCD,''
  JHEP {\bf 0807} (2008) 089
  [arXiv:0803.4192 [hep-th]].
%
\\
  R.~McNees, R.~C.~Myers and A.~Sinha,
  ``On quark masses in holographic QCD,''
  JHEP {\bf 0811} (2008) 056
  [arXiv:0807.5127 [hep-th]].
\\
  P.~C.~Argyres, M.~Edalati, R.~G.~Leigh and J.~F.~Vazquez-Poritz,
  ``Open Wilson Lines and Chiral Condensates in Thermal Holographic QCD,''
  Phys.\ Rev.\ D {\bf 79} (2009) 045022
  [arXiv:0811.4617 [hep-th]].
\\
For other approaches, see:
\\
  R.~Casero, E.~Kiritsis and A.~Paredes,
  ``Chiral symmetry breaking as open string tachyon condensation,''
  Nucl.\ Phys.\ B {\bf 787} (2007) 98
  [hep-th/0702155 [hep-th]].
\\
  K.~Hashimoto, T.~Hirayama and A.~Miwa,
  ``Holographic QCD and pion mass,''
  JHEP {\bf 0706} (2007) 020
  [hep-th/0703024 [hep-th]].
\\
  N.~Evans and E.~Threlfall,
  ``Quark Mass in the Sakai-Sugimoto Model of Chiral Symmetry Breaking,''
  arXiv:0706.3285 [hep-th].
\\
  O.~Bergman, S.~Seki and J.~Sonnenschein,
  ``Quark mass and condensate in HQCD,''
  JHEP {\bf 0712} (2007) 037
  [arXiv:0708.2839 [hep-th]].
\\
  A.~Dhar and P.~Nag,
  ``Sakai-Sugimoto model, Tachyon Condensation and Chiral symmetry Breaking,''
  JHEP {\bf 0801} (2008) 055
  [arXiv:0708.3233 [hep-th]].
\\
  A.~Dhar and P.~Nag,
  ``Tachyon condensation and quark mass in modified Sakai-Sugimoto model,''
  Phys.\ Rev.\ D {\bf 78} (2008) 066021
  [arXiv:0804.4807 [hep-th]].

\bibitem{Wess:1971yu}
  J.~Wess and B.~Zumino,
  ``Consequences of anomalous Ward identities,''
  Phys.\ Lett.\  {\bf 37B} (1971) 95.

\bibitem{Witt2}
  E.~Witten,
  ``Global Aspects of Current Algebra,''
  Nucl.\ Phys.\ B {\bf 223} (1983) 422.

\bibitem{Kaymakcalan:1983qq}
  O.~Kaymakcalan, S.~Rajeev and J.~Schechter,
  ``Nonabelian Anomaly and Vector Meson Decays,''
  Phys.\ Rev.\ D {\bf 30} (1984) 594.

\bibitem{GSW}
  M.~Gell-Mann, D.~Sharp and W.~G.~Wagner,
  ``Decay rates of neutral mesons,''
  Phys.\ Rev.\ Lett.\  {\bf 8} (1962) 261.

\bibitem{SS2}
  T.~Sakai and S.~Sugimoto,
  ``More on a holographic dual of QCD,''
  Prog.\ Theor.\ Phys.\  {\bf 114} (2005) 1083
  [hep-th/0507073].

\bibitem{Skyrme}
  T.~H.~R.~Skyrme,
  Proc.\ Roy.\ Soc.\ Lond.\ A {\bf 260} (1961) 127.
  doi:10.1098/rspa.1961.0018
\\
  T.~H.~R.~Skyrme,
  Proc.\ Roy.\ Soc.\ Lond.\ A {\bf 262} (1961) 237.
  doi:10.1098/rspa.1961.0115
\\
  T.~H.~R.~Skyrme,
  ``A Unified Field Theory of Mesons and Baryons,''
  Nucl.\ Phys.\  {\bf 31} (1962) 556.

\bibitem{AM}
  M.~F.~Atiyah and N.~S.~Manton,
  ``Skyrmions From Instantons,''
  Phys.\ Lett.\ B {\bf 222} (1989) 438.

\bibitem{HM}
  H.~Hata and M.~Murata,
  ``Baryons and the Chern-Simons term in holographic QCD with three flavors,''
  Prog.\ Theor.\ Phys.\  {\bf 119} (2008) 461
  [arXiv:0710.2579 [hep-th]].

\bibitem{HSSY}
  H.~Hata, T.~Sakai, S.~Sugimoto and S.~Yamato,
  ``Baryons from instantons in holographic QCD,''
  Prog.\ Theor.\ Phys.\  {\bf 117} (2007) 1157
  [hep-th/0701280 [hep-th]].

\bibitem{Nair:2005iw}
  V.~P.~Nair,
  ``Quantum field theory: A modern perspective,''
  New York, USA: Springer (2005) 557 p

\bibitem{Witten:1983tx}
  E.~Witten,
  ``Current Algebra, Baryons, and Quark Confinement,''
  Nucl.\ Phys.\ B {\bf 223} (1983) 433.

\bibitem{Guadagnini:1983uv}
  E.~Guadagnini,
  ``Baryons as Solitons and Mass Formulae,''
  Nucl.\ Phys.\ B {\bf 236} (1984) 35.

\bibitem{Mazur:1984yf}
  P.~O.~Mazur, M.~A.~Nowak and M.~Praszalowicz,
  ``SU(3) Extension of the Skyrme Model,''
  Phys.\ Lett.\  {\bf 147B} (1984) 137.

\bibitem{Chemtob:1985ar}
  M.~Chemtob,
  ``Skyrme Model of Baryon Octet and Decuplet,''
  Nucl.\ Phys.\ B {\bf 256} (1985) 600.

\bibitem{Jain:1984gp}
  S.~Jain and S.~R.~Wadia,
  ``Large $N$ Baryons: Collective Coordinates of the Topological Soliton
	in SU(3) Chiral Model,''
  Nucl.\ Phys.\ B {\bf 258} (1985) 713.

\bibitem{Manohar:1984ys}
  A.~V.~Manohar,
  ``Equivalence of the Chiral Soliton and Quark Models in Large N,''
  Nucl.\ Phys.\ B {\bf 248} (1984) 19.

\bibitem{BLR}
  A.~P.~Balachandran, F.~Lizzi, V.~G.~J.~Rodgers and A.~Stern,
  ``Dibaryons as Chiral Solitons,''
  Nucl.\ Phys.\ B {\bf 256} (1985) 525.

\bibitem{Hashimoto:2008zw}
  K.~Hashimoto, T.~Sakai and S.~Sugimoto,
  ``Holographic Baryons: Static Properties and Form Factors from
	Gauge/String Duality,'' 
  Prog.\ Theor.\ Phys.\  {\bf 120} (2008) 1093
  [arXiv:0806.3122 [hep-th]].

\bibitem{Rebhan:2008ur}
  A.~Rebhan, A.~Schmitt and S.~A.~Stricker,
  ``Meson supercurrents and the Meissner effect in the Sakai-Sugimoto model,''
  JHEP {\bf 0905} (2009) 084
  [arXiv:0811.3533 [hep-th]].

\bibitem{Rebhan:2009vc}
  A.~Rebhan, A.~Schmitt and S.~A.~Stricker,
  ``Anomalies and the chiral magnetic effect in the Sakai-Sugimoto model,''
  JHEP {\bf 1001} (2010) 026
  [arXiv:0909.4782 [hep-th]].

\bibitem{Harvey:1996ur}
  J.~A.~Harvey,
  ``Magnetic monopoles, duality and supersymmetry,''
  In *Trieste 1995, High energy physics and cosmology* 66-125
  [hep-th/9603086].

\bibitem{Dijkgraaf:1989pz}
  R.~Dijkgraaf and E.~Witten,
  ``Topological Gauge Theories and Group Cohomology,''
  Commun.\ Math.\ Phys.\  {\bf 129} (1990) 393.

\bibitem{Hashimoto:2009st}
  K.~Hashimoto, N.~Iizuka, T.~Ishii and D.~Kadoh,
  ``Three-flavor quark mass dependence of baryon spectra in holographic QCD,''
  Phys.\ Lett.\ B {\bf 691} (2010) 65
  [arXiv:0910.1179 [hep-th]].

\bibitem{Ishii:2010ib}
  T.~Ishii,
  ``Toward Bound-State Approach to Strangeness in Holographic QCD,''
  Phys.\ Lett.\ B {\bf 695} (2011) 392
  [arXiv:1009.0986 [hep-th]].

\bibitem{Aoki:2012th}
  S.~Aoki, K.~Hashimoto and N.~Iizuka,
  ``Matrix Theory for Baryons: An Overview of Holographic QCD for Nuclear Physics,''
  Rept.\ Prog.\ Phys.\  {\bf 76} (2013) 104301
  [arXiv:1203.5386 [hep-th]].

\bibitem{Matsumoto:2016lxa}
  K.~Matsumoto, Y.~Nakagawa and H.~Suganuma,
  ``A Study of the H-dibaryon in Holographic QCD,''
  arXiv:1610.00475 [hep-th].

\bibitem{Suganuma:2016lmp}
  H.~Suganuma and K.~Matsumoto,
  ``Holographic QCD for H-dibaryon (uuddss),''
  arXiv:1611.05951 [hep-th].

\end{thebibliography}
\end{document}